\documentclass[12pt]{article}
\usepackage{a4wide,amsfonts,float,theorem}

\theorembodyfont{\rmfamily}

\newtheorem{corollary}{Corollary}[section]
\newtheorem{lemma}{Lemma}[section]

\newtheorem{example}{Example}[section]

\floatstyle{ruled}
\restylefloat{table}
\catcode`\@=11
\@addtoreset{equation}{section}
\@addtoreset{table}{section}
\renewcommand{\theequation}{\thesection.\arabic{equation}}

\catcode`\@=12
\def\be{\begin{equation}}
\def\ee{\end{equation}}
\def\ie{{\it i.e.}}
\def\eg{{\it e.g.}}
\def\g{{\cal G}}
\def\R{{\mathbb R}}
\begin{document}
\thispagestyle{empty}
\begin{flushright}
DAMTP 97-49
\\
physics/9706006
\\
May, 1997
\\[0.5cm]
\end{flushright}
\begin{center}
\begin{Large}
{\bf Invariant tensors for simple groups}
\end{Large}
\\[0.4cm]
\renewcommand{\thefootnote}{\fnsymbol{footnote}}
        J. A. de Azc\'arraga\footnote{St. John's College Overseas Visiting
                                     Scholar.
                                  E-mail : j.azcarraga@damtp.cam.ac.uk. On
                                      sabbatical leave from Departamento de
                                      F\'{\i}sica Te\'orica and IFIC
                                      (Centro Mixto Univ. de Valencia-CSIC)
                                      E-46100 Burjassot (Valencia), Spain.},
        A. J. Macfarlane\footnote{E-mail : a.j.macfarlane@damtp.cam.ac.uk},
        A. J. Mountain\footnote{E-mail : a.j.mountain@damtp.cam.ac.uk} and
        J. C. P\'erez Bueno\footnote{E-mail : pbueno@lie.ific.uv.es. On leave
                                     of absence from Departamento de
                                     F\'{\i}sica Te\'orica and IFIC
                                     (Centro Mixto Univ. de Valencia-CSIC)
                                     E-46100 Burjassot (Valencia), Spain.}
\\[0.3cm]
\setcounter{footnote}{0}
\begin{it}
Department of Applied Mathematics and Theoretical Physics, \\
Silver St., Cambridge CB3 9EW, UK
\end{it}
\\[0.3cm]
\end{center}

\begin{abstract}
  The forms of the invariant primitive tensors for the simple Lie
  algebras $A_l, B_l, C_l$ and $D_l$ are investigated.  A new family
  of symmetric invariant tensors is introduced using the non-trivial
  cocycles for the Lie algebra cohomology. For the $A_l$ algebra it is
  explicitly shown that the generic forms of these tensors become zero
  except for the $l$ primitive ones and that they give rise to the
  $l$ primitive Casimir operators. Some recurrence and duality
  relations are given for the Lie algebra cocycles. Tables for the
  3- and 5-cocycles for $su(3)$ and $su(4)$ are also provided. Finally, new
  relations involving the $d$ and $f$ $su(n)$ tensors are given.
\end{abstract}

\section{Introduction}

We devote this paper to a systematic study of the symmetric and
skewsymmetric primitive invariant tensors that may be constructed on a
compact simple Lie algebra $\g$.  The symmetric invariant tensors give
rise to the Casimirs of $\g$; the skewsymmetric ones determine the
non-trivial cocycles for the Lie algebra cohomology (see {\it e.g.}
\cite{CE}).  It is well known \cite{RACAH,GEL,AK,GR,LCB,PP,OKPA,SOK}
that there are $l$ such invariant symmetric primitive polynomials of
order $m_i$ ($i=1,\dots,l=\mbox{rank\ of\ }\g$), which determine $l$
independent primitive Casimir operators of the same order, as well as
$l$ skewsymmetric invariant primitive tensors $\Omega^{(2m_i-1)}$ of
order ($2m_i-1$). The latter determine the non-trivial cocycles for
the Lie algebra cohomology, their order being related to the
topological properties of the associated compact group manifold $G$
which, from the point of view of the real homology, behaves as
products of $l$ $S^{(2m_i-1)}$ spheres
\cite{CAR,PON,HOPF,HOD,SAMEL,BOREL,BOTT,LJB,HIGHER}.  The
lowest examples of these tensors/polynomials ($m_1=2$) are the Killing
tensor (which is a multiple of $\delta_{ij}$ for a compact algebra),
the quadratic Casimir operator and the fully skewsymmetric structure
constants of the simple algebra $\g$, which determine a three-cocycle
on $\g$ (see Example \ref{ex1} below). The study of the invariant
primitive tensors on $\g$ (and especially in the most interesting case
$\g=su(n)$) is not only mathematically relevant; their properties
determine essential aspects of the physical theories based on the
associated group $G$. These range from the form of the vertices in
Feynman diagrams to the presence or absence of non-abelian anomalies
in gauge theories (see in this last respect the articles in
\cite{TREIMAN} and references therein; see also \cite{AI}).  They are
also relevant in other instances as {\it e.g.}, in (higher order)
Yang-Mills duality problems \cite{BB}, WZWN terms (see, {\it e.g.}
\cite{AIM} and references therein), ${\cal W}$-symmetry in conformal
field theory \cite{BOUSCHOU}, the construction of effective
actions \cite{DHW}, etc. It is then convenient to have an explicit
expression for the primitive tensors of the different orders as well
as a convenient basis for the vector spaces of the invariant tensors
of a given order.

The simplest way of obtaining an invariant symmetric tensor
$k^{(r)}$ on $\g$ is by computing the trace
of the symmetrised  product of $r$ generators.
This symmetric trace (which may give a zero result depending on $r$ and
on the specific algebra $\g$ being considered) will not vanish, however,
for arbitrary $r$. As a result, the trace will give rise to arbitrarily
higher order tensors that cannot  be primitive and
independent of those of lower order $m_i$ ($i=1,\dots,l$), and the same
will apply to the Casimir operators constructed from them.
This indicates that it is convenient to introduce a new family of tensors
which is free of this problem. We shall do this in Sec.
\ref{invsec3} by introducing a new family of symmetric invariant tensors
$t^{(m_i)}$ from the $l$ primitive $(2m_i-1)$-cocycles $\Omega^{(2m_i-1)}$,
($i=1,\dots,l$). These tensors will turn out to be `orthogonal' in a
precise sense (see Lemma \ref{lem3.3}).
We shall also show how the $l$ primitive
Casimirs may be equivalently obtained
from the $t^{(m_i)}$ tensors or from the cocycles $\Omega^{(2m_i-1)}$.

  For the $su(n)$ algebras many results and techniques are already available.
For instance, we can construct recursively \cite{SUDBERY}
(see also \cite{BBSS,AK}) the so-called
$d$-family of symmetric invariant tensors of order $m$ starting from the
symmetric $d_{ijk}$ (for $su(n), n>2$) and symmetrising the result
$d_{(i_1\dots i_m)}$.
This family (see Sec. \ref{suse6.1}) has a status like the $k$
family \ie, for $n$ fixed and $m$ large enough the $d^{(m)}$ tensors may be
expressed as linear combinations of products of lower order ones.
For fixed $m$ and $n$ (in $su(n)$) sufficiently
large (in fact, for $n\ge m$)
it is known how to define a basis of the vector spaces ${\cal V}^{(m)}$ of
invariant symmetric tensors of order $m$, and via known identities for the
$d^{(m)}$, how this basis reduces when $n<m$.
We use these properties in our discussion of the $t^{(m_i)}$ family.
In particular we see that a formal attempt to construct $t$-tensors of
higher rank than is allowed from their definition necessarily yields an
identically vanishing result (Sec.\ref{suse6.2}).
In proving this, we have needed to extend the set of identities
for the $d$ and $f$ $su(n)$ tensors in the literature known to us.

The paper is organised as follows. After a short general discussion of
the invariance properties of tensors and Casimir operators on $\g$ in
Sec.\ref{invsec2} the expression of the ($2m_i-1$)-cocycles is given
in Sec.\ref{invsec3}, where the $t^{(m_i)}$ family of invariant
symmetric tensors is introduced. Sec.\ref{cocycles} illustrates these
general considerations for the $su(3)$ and $su(4)$ algebras.
Sec.\ref{se4} provides a general discussion of the primitivity of the
invariant symmetric tensors and Casimir operators for the four
infinite series $A_l, B_l, C_l$ and $D_l$ and shows explicitly, if not
systematically, how a given primitive polynomial becomes algebraically
dependent (non-primitive) when the rank of the algebra $\g$ is reduced
sufficiently.  Due to the special relevance of the $su(n)$ algebras,
Sec.\ref{invsym} is devoted to illustrate these ideas for the $A_l$
case and, in particular, the usefulness of the $t^{(m_i)}$ family of
tensors.  Sec.\ref{invsec7} discusses, again for the four infinite
series, the properties of the $(2m_i-1)$-cocycles.  The topological
properties underlying the compact group manifolds provide a clue to
establish, using the Hodge star $*$ operator, duality properties for
the Lie algebra cocycles. This is done in Sec.\ref{invsec8}, and some
of the formulae are illustrated by using the explicit
results in Sec.\ref{cocycles}.
Finally an Appendix develops some properties of
the $d$ and $f$ $su(n)$ tensors for arbitrary $n$, and collects a
number of new expressions which are needed for the derivation of
crucial results in the main text.

\section{Invariant symmetric polynomials and Casimir operators}
\label{invsec2}

Let $\g$ be a simple algebra of rank $l$ with basis
$\{X_{i}\}$, $[X_i,X_j]=C_{ij}^k X_k$, $i=1,\dots,r=\mbox{dim}\,\g$,
and let $G$ be its (compact) associated Lie group\footnote{In the
general discussions we adopt the `mathematical' convention and
take antihermitian generators
$X_i$ and hence negative definite Killing tensor
$K_{ij}=\mbox{Tr}(adX_i adX_j)\propto -\delta_{ij}$ since
$\g$ is compact. When we consider explicit $su(n)$ examples, we
follow the `physical' convention and use hermitian generators
$T_i, X_i=-iT_i$. In this case, an $i$ accompanies
the $C_{ij}^k$ in the $r.h.s$ of the commutators.
When the generators are assumed to be
in {\it matrix} form, we take them in the defining representation
of the algebra. We  always use unit
metric and hence there is no distinction among
upper and lower indices, their position
being dictated by notational convenience.}.
Let $\{\omega^{j}\}$ be the dual basis in $\g^*$, $\omega^{j}(X_i)=\delta_i^j$,
and consider a $G$-invariant symmetric tensor $h$ of order $m$
\be
h=h_{i_1\dots i_m} \omega^{i_1}\otimes\dots\otimes\omega^{i_m} \quad.
\label{seI0}
\ee
The $G$-invariance of $h$ means that
\be
\sum_{s=1}^m C^\rho_{\nu i_s} h_{i_1\dots \widehat{i_s}\rho i_{s+1} \dots
i_m}=0
\quad.
\label{seIii}
\ee
This is the case of the symmetric tensors $k^{(m)}$ given by the
coordinates\footnote{
\label{foot2}
Indices inside round brackets $(i_1,\dots,i_m)$ will always be
understood as symmetrised with unit weight \ie, with a factor
$1/m!$. We use the same unit weight convention to antisymmetrise
indices inside square brackets  $[i_1,\dots,i_m]$.}
\be
k_{i_1\dots i_m}= {1\over m!}\mbox{sTr}(X_{i_1}\dots X_{i_m})
\equiv \mbox{Tr}(X_{(i_1}\dots X_{i_m)})\quad,
\label{seIi}
\ee
where sTr is the symmetric trace,
$\displaystyle \mbox{sTr}(X_{i_1}\dots X_{i_m})=
\sum_{\sigma\in S_m} \mbox{Tr}(X_{i_{\sigma(1)}}\dots X_{i_{\sigma(m)}})$,
which is clearly ad-invariant\footnote{
We denote by $k^{(m)}$ the
invariant symmetric tensors coming from the
symmetric trace (\ref{seIi}). In fact (see \cite{GR})
a complete set of $l$ primitive (see below) invariant
tensors may be constructed in this way
by selecting suitable representations. Other families of
symmetric invariant tensors will be identified by an appropriate
letter, ({\it e.g.} $t$, $d$, $v$). Generic symmetric invariant
polynomials are denoted by $h$.}.
In particular,
\be
k_{ij}=\mbox{Tr}(X_i X_j)=\kappa\delta_{ij}
\label{basickilling}
\ee
where, for instance, $\kappa=-1/2$ for the generators $X_i$ of the
defining representation of $su(n)$.

Since $\g\sim T_e(G)$, the tangent space at the identity of $G$,
we may use a left translation $L_g,\ g\in G$, to obtain a
left-invariant (LI) $m$-tensor $h(g)$ on the group manifold $G$ from the
$m$-linear mapping $h:\g\times\mathop{\cdots}\limits^m\times\g\to\R$.
Its expression is the same as (\ref{seI0}) where now the $\{\omega^j\}$ are
replaced by the LI one-forms $\omega^i (g)$ on
$G$, and (\ref{seIii}) now follows from the fact that (see, {\it e.g.}
\cite{AI})
\be
L_{X_\nu(g)}\omega^\rho (g) =-C^\rho_{\nu i}\omega^ i (g)
\quad,\quad \nu,\rho,i=1,\dots,r \quad,
\label{seIiii}
\ee
where $L_{X_\nu (g)}$ is the Lie derivative with respect to the vector field
$X_\nu(g)$ obtained by applying the (tangent) left translation
map $L_g^T$ to $X_\nu(e)=X_\nu$;
clearly the duality relation is maintained for the vector fields $\{X_i(g)\}$
and one-forms $\{\omega^j (g)\}$. In this context, the $G$-invariance
condition (\ref{seIii}) reads $L_{X_{\nu}(g)}h(g)=0$.

Let $G$ moreover be compact so that the Killing tensor may be taken as
the unit matrix and let $h_{i_1\dots i_m}$ be an arbitrary symmetric
invariant tensor.  Then the order $m$ element in the enveloping
algebra ${\cal U}(\g)$ defined by \be {\cal C}^{(m)}=h^{i_1\dots i_m}
X_{i_1}\dots X_{i_m}
\label{seIiv}
\ee
commutes with all elements in $\g$.
This is so because the commutator $[X_\rho,{\cal C}^{(m)}]$ may be written as
\be
[X_\rho,{\cal C}^{(m)}]=
\sum_{s=1}^m C_{\rho \nu } ^{i_s} h^{i_1\dots \widehat{i_s} \nu\dots  i_m}
X_{i_1}\dots X_{i_m} =0
\quad,
\label{seIv}
\ee
which is indeed zero as a result of the invariance condition (\ref{seIii}).
In fact, the only conditions for the $m$-tensor $h$ to generate
a Casimir operator ${\cal C}^{(m)}$ of $\g$ of order $m$ are its
symmetry (non-symmetric indices would allow us to reduce the order $m$ of
${\cal C}^{(m)}$
by replacing certain products of generators by commutators) and its invariance
(eq. (\ref{seIv}));
$h$ does not need to be obtained from a symmetric trace (\ref{seIi}).
This leads to

\begin{lemma}
({\it Casimirs and $G$-invariant symmetric polynomials})
\\
Let $h$ be an invariant symmetric tensor of order $m$. Then,
${\cal C}^{(m)}=h^{i_1\dots i_m} X_{i_1}\dots X_{i_m}$ is a Casimir of
$\g$ of the same order $m$.
\label{lem2.1}
\end{lemma}

It is well known \cite{RACAH,GEL,LCB,AK,GR,PP,OKPA,SOK}
that a simple algebra of rank $l$ has $l$ independent
(primitive) Casimir-Racah operators of order $m_1,\dots,m_l$, the first of
them given by the standard Casimir \cite{CASIMIR} operator $K_{ij} X^i X^j$
obtained from the Killing tensor $(m_1=2)$
Thus,  there must be (Cayley-Hamilton) relations among the invariant
tensors obtained from (\ref{seIi}) for $m>m_l$ or otherwise one would
obtain an arbitrary number of primitive Casimirs. We shall study this
problem in Sec.\ref{se4} and apply our results to the $su(n)$ algebras in
Sec.\ref{invsym}.

\section{Invariant skewsymmetric tensors and cocycles}
\label{invsec3}

Let $\theta(g)=\omega^i(g) X_i$ be the LI canonical form on a simple and
compact group $G$, and consider the $q$-form
Tr$(\theta\wedge\mathop{\cdots}\limits^q\wedge\theta)$.
Due to the cyclic property of the trace and the anticommutativity of one-forms,
this form is zero for $q$ even.
Let $q$ be odd. Then,
\be
\Omega^{(q)}(g)={1\over q!} \mbox{Tr}
(\theta\wedge\mathop{\cdots}\limits^q\wedge\theta)
\label{seIIi}
\ee
is a closed form on the group manifold $G$, since
$d\Omega\propto \mbox{Tr}
(\theta\wedge\mathop{\cdots}\limits^{q+1}\wedge\theta) =0$ on account of the
Maurer-Cartan equations
$d\theta = -\theta\wedge\theta$.
Since $\Omega(g)$ is not exact (it cannot be the exterior differential of the
$(q-1)$-form $\mbox{Tr}(\theta\wedge\mathop{\cdots}\limits^{q-1}\wedge\theta)$
which is zero because $q-1$ is even) it defines a Chevalley-Eilenberg
\cite{CE} Lie algebra $q$-cocycle.
If we set $q=2m-1$, we find that
\be
\begin{array}{rl}
\Omega^{(2m-1)}(g)
&
\displaystyle
= {1\over (2m-1)!} \mbox{Tr} (X_{i_1}\dots X_{i_{2m-1}})
\omega^{i_1}(g)\wedge\dots\wedge\omega^{i_{2m-1}}(g)
\\[0.3cm]
&
\displaystyle
= {1\over (2m-1)!} {1 \over 2^{m-1} } \mbox{Tr}(
[X_{i_1},X_{i_2}][X_{i_3},X_{i_4}]\dots [X_{i_{2m-3}},X_{i_{2m-2}}]X_{i_{2m-1}}
)
\\[0.4cm]
&
\displaystyle\qquad\qquad\qquad
\cdot
\omega^{i_1}(g)\wedge\dots\wedge\omega^{i_{2m-1}}(g)
\\[0.4cm]
&
\displaystyle
= {1\over (2m-1)!} {1 \over 2^{m-1} }
C_{i_1 i_2}^{l_1}\dots C_{i_{2m-3} i_{2m-2}}^{l_{m-1}}
\mbox{Tr}( X_{l_1}\dots X_{l_{m-1}} X_\sigma )
\\[0.3cm]
&
\displaystyle\qquad\qquad\qquad
\cdot
\omega^{i_1}(g)\wedge\dots\wedge\omega^{i_{2m-2}}(g) \wedge \omega^\sigma(g)
\quad.
\end{array}
\label{seIIii}
\ee
The trace may in fact be replaced by a unit weight symmetric trace \ie,
by Tr $\sim (1/m!)$sTr in (\ref{seIIii}), since the skewsymmetry in
$i_1\dots i_{2m-2}$ results in symmetry in $l_1\dots l_{m-1}$.
The factor ${1\over 2^{m-1}}$ is unimportant\footnote{
\label{foot3} We recall that
the cohomology space is a vector space, and that numerical
factors, although they relate {\it inequivalent} cocycles (different vectors
in a given cohomology space $H^{(2m-1)}(\g,\R)$), are unimportant here;
they determine only the normalisation of the different tensors.}
and will be ignored from now on.
Hence, we may define the $(2m-1)$-form on $G$ representing a Lie algebra
$(2m-1)$-cocycle by
\be
\Omega^{(2m-1)}(g) = {1\over (2m-1)!}
\Omega_{i_1\dots i_{2m-2} \sigma}
\omega^{i_1}(g)\wedge\dots\wedge\omega^{i_{2m-2}}(g) \wedge \omega^\sigma(g)
\quad,
\label{seIIiii}
\ee
the coordinates of which are given by the constant skewsymmetric tensor
\be
\Omega_{i_1\dots i_{2m-2} \sigma} =
{1\over (2m-1)!}
\epsilon^{j_1\dots j_{2m-2} \rho}_{i_1\dots i_{2m-2} \sigma}
C_{j_1 j_2}^{l_1}\dots C_{j_{2m-3} j_{2m-2}}^{l_{m-1}}
k_{l_1\dots l_{m-1} \rho}
\quad,
\label{seIIiv}
\ee
where $k_{l_1\dots l_{m-1} \rho}$ is the symmetric tensor of
(\ref{seIi})\footnote{
The invariance properties of (\ref{seIIiv}) follow from the
invariance of $k_{l_1\dots l_{m-1} \rho}$ and thus do not depend
on the fact that it is expressed by a symmetric trace (\ref{seIi}).}
and the $\epsilon$ tensor is defined by
\be
\epsilon_{\alpha_1 \dots \alpha_n}^{\beta_1\dots\beta_n}
=\sum_{\sigma\in S_n}
(-1)^{\pi(\sigma)}
\delta_{\alpha_1}^{\beta_{\sigma(1)}}\dots
\delta_{\alpha_n}^{\beta_{\sigma (n)}}
\quad,
\label{seIIaiii}
\ee
where $\pi (\sigma)$ is the parity of the permutation $\sigma$.
Although (\ref{seIIiv}) is what follows naturally from (\ref{seIIi})
(setting aside the ignored factor $1 / 2^{(m-1)}$) it is
convenient to notice that part of the antisymmetrisation carried out by
$\epsilon^{j_1\dots j_{2m-2} \rho}_{i_1\dots i_{2m-2} \sigma}$ is unnecessary,
since (cf. (\ref{seIIiv})) may be rewritten as

\be
\begin{array}{rl}
\Omega_{\rho i_2\dots i_{2m-2} \sigma} & =
\displaystyle
{1\over {(2m-3)!}}
\epsilon^{j_2\dots j_{2m-2}}_{i_2\dots i_{2m-2}}
C_{\rho j_2}^{l_1}\dots C_{j_{2m-3} j_{2m-2}}^{l_{m-1}}
k_{l_1\dots l_{m-1} \sigma}
\\[0.3cm]
&
\displaystyle
\equiv C_{\rho [i_2}^{l_1}\dots C_{i_{2m-3} i_{2m-2}]}^{l_{m-1}}
k_{l_1\dots l_{m-1} \sigma}
\end{array}
\label{seIIv}
\ee
due to the skewsymmetry in $\rho$ and $\sigma$ of its $r.h.s$. This
follows from the invariance of the symmetric polynomial
$k_{l_1\dots l_{m-1} \sigma}$ and is a generalisation
of the simple $m=2$ case for which (\ref{seIIv}) gives
\be
\Omega_{\rho j \sigma}
=k ([X_\rho,X_j],X_\sigma)=
k(X_\rho,[X_j,X_\sigma]) = - k ([X_\sigma,X_j],X_\rho)
= -\Omega_{\sigma j \rho}\quad.
\ee
Indeed, for an arbitrary invariant symmetric tensor $h$ on
$\g$ of order $m$ we have from (\ref{seIIv})
$$
\begin{array}{l}
\displaystyle
(2m-3)!\,\Omega_{\rho i_2\dots i_{2m-2} \sigma}=
\epsilon^{j_2\dots j_{2m-2}}_{i_2\dots i_{2m-2}}
h([X_\rho,X_{j_2}],[X_{j_3},X_{j_4}],\dots,[X_{j_{2m-3}},X_{j_{2m-2}}],
X_\sigma)
\\[0.3cm]
\displaystyle
=
-\epsilon^{j_2\dots j_{2m-2}}_{i_2\dots i_{2m-2}}
\sum_{s=2}^{m-1}
h(X_\rho,[X_{j_3},X_{j_4}],\dots,[[X_{j_{2s-1}},X_{j_{2s}}],X_{j_{2}}],
\dots,[X_{j_{2m-3}},X_{j_{2m-2}}],X_\sigma)
\\[0.3cm]
\displaystyle
-\epsilon^{j_2\dots j_{2m-2}}_{i_2\dots i_{2m-2}}
h(X_\rho,[X_{j_3},X_{j_4}],\dots,[X_{j_{2m-3}},X_{j_{2m-2}}],[X_\sigma,
X_{j_2}])
\\[0.3cm]
\displaystyle
=
\epsilon^{j_2\dots j_{2m-2}}_{i_2\dots i_{2m-2}}
h(X_\rho,[X_{j_3},X_{j_4}],\dots,[X_{j_{2m-3}},X_{j_{2m-2}}],[X_{j_2},
X_\sigma])
\\[0.3cm]
\displaystyle
=
-\epsilon^{j_2\dots j_{2m-2}}_{i_2\dots i_{2m-2}}
h([X_\sigma,X_{j_2}],[X_{j_3},X_{j_4}],\dots,[X_{j_{2m-3}},X_{j_{2m-2}}],
X_\rho)
=-(2m-3)! \,\Omega_{\sigma i_2\dots i_{2m-2} \rho \quad,}
\end{array}
$$
where we have used the invariance of $h$ in the second equality, the
Jacobi identity in the third (to see that every term in the summation
symbol is zero) and the symmetry of $h$ in the fourth one. The fact that
the skewsymmetric tensors $\Omega^{(2m_i-1)}$ expressed by their
coordinates (\ref{seIIiv}) or (\ref{seIIv}) are indeed
($2m_i-1$)-cocycles follows from the Chevalley-Eilenberg approach to
Lie algebra cohomology already mentioned \cite{CE}; for a direct proof
which uses only the symmetry and invariance properties of the tensor
used in its definition (\ref{seIIv}) see \cite{APPB}.
\begin{example}
Let $m=2$.
Using $\delta_{ij}$ (rather than $k_{ij}$) as the lowest
order invariant polynomial,
(\ref{seIIv}) gives
\be
\Omega_{i_1 i_2 \sigma}=C^{l_1}_{i_1 i_2} \delta_{l_1 \sigma} = C_{i_1 i_2
\sigma}
\label{seIIvii}
\ee
\ie, the three-cocycle $\Omega_{i_1 i_2 i_3}$ is determined by the structure
constants of $\g$. It follows that the $3rd$ Lie algebra cohomology group
$H^3(\g)$ is non-zero for $\g$ simple, as is well known.
\label{ex1}
\end{example}

\begin{example}
Let $k_{i_1 i_2 i_3}$ be a $3rd$-order invariant symmetric polynomial.
Such a $3rd$-order
polynomial exists only for $su(n)$, $n>2$ (this is the reason why in four
dimensions only these groups are unsafe for non-abelian anomalies).
Then, the $su(n)$ five-cocycle is given by (\ref{seIIv})
\be
\Omega_{\rho i_1 i_2 i_3 \sigma} ={1\over {3!}}
\epsilon^{j_1 j_2 j_3}_{i_1 i_2 i_3}
C^{l_1}_{\rho j_1} C^{l_2}_{j_2 j_3} k_{l_1 l_2 \sigma}
\quad.
\label{seIIviii}
\ee
The coordinates of $\Omega^{(3)}$, $\Omega^{(5)}$ for $su(3)$ and
for $su(4)$ are given in tables \ref{table1},
\ref{table2} and \ref{table4}, \ref{table6} respectively.
The expression of the three-cocycle follows directly from
the structure constants, eq. (\ref{seIIvii}); for five-cocycles
we have used the symmetric Gell-Mann tensor $d_{ijk}$
in (\ref{seIIviii}) rather than $k_{ijk}$
[$k_{ijk}=(1/3!)(-i/2)^3\mbox{sTr}(\lambda_i\lambda_j\lambda_k)$
=$(i/4)d_{ijk}$, see (\ref{seIIai}) below].
\end{example}

Since the $(2m-1)$-forms $\Omega^{(2m-1)}(g)$ (\ref{seIIiii}) are invariant,
the coordinates of any $(2m-1)$-cocycle on the simple Lie algebra also satisfy
the relation (cf. (\ref{seIii}))
\be
\sum_{s=1}^{2m-1} C^\rho_{\nu i_s}
\Omega_{i_1\dots\widehat{i_s}\rho i_{s+1} \dots i_{2m-1}}=0
\quad.
\label{seIIix}
\ee
The simplest case corresponds to Example \ref{ex1} for which
\be
C^\rho_{\nu i_1} C_{\rho i_2 i_3} +
C^\rho_{\nu i_2} C_{i_1 \rho i_3} +
C^\rho_{\nu i_3} C_{i_1 i_2 \rho} = 0
\label{seIIx}
\ee
is the Jacobi identity
$C^\rho_{\nu [i_1} C_{i_2 i_3]\rho}=0$.
Similarly, the five-cocycle (which only exists for $su(n)$, $n>2$) satisfies
\be
C^\rho_{\nu i_1} \Omega^{(5)}_{\rho i_2 i_3 i_4 i_5} +
C^\rho_{\nu i_2} \Omega^{(5)}_{i_1 \rho i_3 i_4 i_5} +
C^\rho_{\nu i_3} \Omega^{(5)}_{i_1 i_2 \rho i_4 i_5} +
C^\rho_{\nu i_4} \Omega^{(5)}_{i_1 i_2 i_3 \rho i_5} +
C^\rho_{\nu i_5} \Omega^{(5)}_{i_1 i_2 i_3 i_4 \rho}
=0\quad.
\label{seIIxi}
\ee

\begin{lemma}\
\\
Let $h_{l_1\dots l_m}$ be a symmetric $G$-invariant polynomial. Then,
\be
\epsilon^{j_1 \dots j_{2m}}_{i_1\dots i_{2m}}
C^{l_1}_{j_1 j_2}\ldots C^{l_m}_{j_{2m-1} j_{2m}}
h_{l_1\dots l_m}=0 \quad.
\label{simplelemma}
\ee
\label{lem3.1}
\end{lemma}
{\it Proof:}\quad
By replacing $C^{l_m}_{j_{2m-1} j_{2m}}h_{l_1\dots lm}$ in the $l.h.s$ of
(\ref{simplelemma}) by the other terms in (\ref{seIii}) we get
$$
\epsilon^{j_1 \dots j_{2m}}_{i_1 \dots i_{2m}}
C^{l_1}_{j_1 j_2}\ldots C^{l_{m-1}}_{j_{2m-3} j_{2m-2}}
(\sum_{s=1}^{m-1}C^{k}_{j_{2m-1} l_s} h_{l_1\ldots l_{s-1} k l_{s+1}\ldots
l_{m-1}\,j_{2m}})
\quad,
$$
which vanishes since all terms in the sum include products of the form
$C^s_{jj'}C^k_{sj''}$ antisymmetrised in $j,j',j''$, which are zero due to
the Jacobi identity, {\it q.e.d.}
Note that if $h_{l_1\dots l_m}$ is identified with
$k_{l_1\dots l_m}$ in (\ref{seIi}) the equality follows immediately
from the fact that $l.h.s.$ (\ref{simplelemma}) are the coordinates
of $\Omega_{i_1\dots i_{2m}}$, and $\Omega^{(2m)}\propto
\mbox{Tr}(\theta\wedge \mathop{\cdots}\limits^{2m}\wedge\theta)=0$.

\begin{corollary}
\ \\
Let $h$ now be a non-primitive symmetric $G$-invariant polynomial,
\ie, such that its coordinates are given by
\be
h_{i_1\dots i_p j_1\dots j_q} =
h^{(p)}_{(i_1\dots i_p}  h^{(q)}_{j_1\dots j_q)}
\quad,
\label{nonprim}
\ee
where in the $r.h.s.$ $(\dots)$ indicates unit weight
symmetrisation (see footnote \ref{foot2})
and $h^{(p)}$ and $h^{(q)}$ are symmetric invariant polynomials.
Then the cocycle $\Omega^{2(p+q)-1}$ associated to (\ref{nonprim}) by
(\ref{seIIiv}) is zero.
\label{cor3.1}
\end{corollary}
{\it Proof:}\quad
$\Omega^{2(p+q)-1}$ is given by
$$
\begin{array}{l}
\displaystyle
\Omega^{2(p+q)-1}_{i_1\dots i_{2(p+q)-1}}={1\over{(2(p+q)-1)!}}
\epsilon_{i_1\dots i_{2(p+q)-1}}^{j_1\dots j_{2(p+q)-1}}
C^{l_1}_{j_1 j_2}\dots C^{l_p}_{j_{2p-1} j_{2p}}
C^{m_1}_{j_{2p+1} j_{2p+2}} \dots C^{m_{q-1}}_{j_{2(p+q)-3} j_{2(p+q)-2}}
\\[0.3cm]
\qquad \qquad \qquad h^{(p)}_{(l_1\dots l_p}
h^{(q)}_{m_1\dots m_{q-1} j_{2(p+q)-1} ) }
\end{array}
$$
and is zero by virtue of (\ref{simplelemma}), {\it q.e.d.}

\medskip
\noindent
By definition, primitive tensors are not the product of lower order
tensors, but may contain non-primitive terms.  It follows from
Corollary \ref{cor3.1} that only the primitive term in $k$ contributes
to the cocycle (\ref{seIIv}). As a result, different families of
symmetric tensors differing in non-primitive terms lead to
proportional cocycles. Thus, and as far as the construction of the
cocycles is concerned, we may use any family $h^{(m)}$ of symmetric
invariant primitive tensors in (\ref{seIIv}), not necessarily that
given by (\ref{seIi}); for instance, for $su(n)$ we may use the $d$
family of Sec \ref{suse6.1}. We shall use (\ref{seIIv}) with the
definition (\ref{seIi}) unless otherwise indicated.

We introduce now a new type of symmetric invariant polynomials by
using the cocycles. Their interest will be made explicit in Sec. \ref{invsym}.

\begin{lemma}
({\it Invariant symmetric polynomials from primitive cocycles})
\\
Let $\Omega^{(2m-1)}$ be a primitive cocycle.
The $l$ polynomials $t^{(m)}$ given by
\be
{t}^{i_1 \dots i_{m}}=[\Omega^{(2m-1)}]^{j_1 \dots j_{2m-2} i_m}
C^{i_1}_{j_1 j_2} \dots C^{i_{m-1}}_{j_{2m-3} j_{2m-2}}
\label{seIIxii}
\ee
are $G$-invariant, symmetric and
primitive. Moreover, they are traceless
for $m>2$.
\label{lem3.2}
\end{lemma}
{\it Proof:}\quad
By construction ${t}^{i_1 \dots i_{m}}$
is an invariant polynomial (it is
obtained by contracting the invariant tensors $C$ and $\Omega$).
It follows from (\ref{seIIxii}) that it is symmetric under interchange of the
$(i_1 \dots i_{m-1})$ indices.
Now
$$
\begin{array}{rl}
{t}^{i_1 \dots i_{m}}=&
[\Omega^{(2m-1)}]^{j_1 \dots j_{2m-2} i_m}
C^{i_1}_{j_1 j_2} C^{i_2}_{j_3 j_4} \dots C^{i_{m-1}}_{j_{2m-3} j_{2m-2}}
\\[0.3cm]
= & \displaystyle
-\sum_{s=2}^{2m-2}
([\Omega^{(2m-1)}]^{i_1 j_2 \dots \widehat{j_s} \rho \dots j_{2m-2} i_m}
C^{j_s}_{\rho j_2})C^{i_2}_{j_3 j_4} \dots C^{i_{m-1}}_{j_{2m-3} j_{2m-2}}
\\[0.3cm]
& \displaystyle
-[\Omega^{(2m-1)}]^{i_1 j_2 \dots j_{2m-2} \rho}
C^{i_m}_{\rho j_2}C^{i_2}_{j_3 j_4} \dots C^{i_{m-1}}_{j_{2m-3} j_{2m-2}}
\\[0.3cm]
=
&
-[\Omega^{(2m-1)}]^{i_1 j_2 \dots j_{2m-2} \rho}
C^{i_m}_{\rho j_2}C^{i_2}_{j_3 j_4} \dots C^{i_{m-1}}_{j_{2m-3} j_{2m-2}}
\\[0.3cm]
=
&
[\Omega^{(2m-1)}]^{\rho j_2 \dots j_{2m-2} i_1}
C^{i_m}_{\rho j_2}C^{i_2}_{j_3 j_4} \dots C^{i_{m-1}}_{j_{2m-3} j_{2m-2}}
\\[0.3cm]
=
&
{t}^{i_m i_2 \dots i_{m-1} i_{1}} \quad,
\end{array}
$$
where we have used the invariance (\ref{seIIix}) of
$\Omega^{(2m-1)}$ in the second equality, the Jacobi identity in the
third and the skewsymmetry of $\Omega^{(2m-1)}$ in the fourth.  Thus
${t}$ is invariant under the change $i_1\leftrightarrow i_m$ and hence
it is a completely symmetric tensor. For $m=2$ eq.  (\ref{seIIxii}) is
proportional to the unit matrix since
\be
C^{j_1 j_2 i_2} C^{i_1}_{j_1 j_2}=K^{i_1 i_2}
\label{seIIxiiaa}
\ee
is the Killing tensor $K$.
If we contract (\ref{seIIxii}) with $\delta_{i_1 i_2}$ for $m>2$ we see that
\be
{{t}_\sigma}^{\sigma i_3 \dots i_m} =0
\label{tracelessness}
\ee
by using the Jacobi identity for $C_{\sigma  j_1 j_2} C^{\sigma}_{j_3 j_4}$,
{\it q.e.d.}

Since $k_{i_1 i_2} \propto \delta_{i_1 i_2}$,
the tracelessness of (\ref{seIIxii}) may be seen as ${t}^{i_1 \dots i_m}$
having zero contraction with $k_{ij}$.
This extends to the full contractions with all higher order
symmetric invariant tensors by means of the following

\begin{lemma}
\
\\
Let ${t}^{i_1 \dots i_m}$ be the symmetric invariant polynomial given by
(\ref{seIIxii}).
Then
\be
{t}^{i_1 \dots i_l i_{l+1} \dots i_m} {t}_{i_1 \dots i_l}=0
\quad, \quad \forall l<m
\quad.
\label{seIIxiia}
\ee
\label{lem3.3}
\end{lemma}
{\it Proof:}
(\ref{seIIxiia}) is a consequence of Lemma \ref{lem3.1}, {\it q.e.d.}

\medskip
\noindent
{\it Note.} Eq. (\ref{seIIxiia}) implies the `orthogonality' of the
different polynomials ${t}^{(m)}_{i_1\dots i_m}$ obtained from
$(2m-1)$-cocycles.  Thus, a basis for the space ${\cal V}^{(m)}$ of
invariant symmetric polynomials of order $m$ is given by the
symmetrised products of the primitive symmetric invariant polynomials
(or their powers) leading to a symmetric invariant tensor of order
$m$. We may express this as a

\begin{corollary}
\
\\
The symmetric invariant primitive polynomial
$t^{(m)}$ and the symmetrised products $t^{(m-r_1)}\otimes t^{(r_1)}$,
$t^{(m-r_1 -r_2)}\otimes t^{(r_1)}\otimes t^{(r_2)}$ (see (\ref{nonprim}))
etc., constitute a basis of the vector space ${\cal V}^{(m)}$ of the
$G$-invariant symmetric polynomials on ${\cal G}$ of order $m$.
\label{corbasis}
\end{corollary}

\begin{example}
For $su(n)$ the vector space of symmetric invariant
polynomials of order six is given by
${t}^{(6)}_{i_1\dots i_6}$, ${t}^{(4)}_{(i_1\dots i_4}\delta_{i_5 i_6)}$,
${t}^{(3)}_{(i_1 i_2 i_3} {t}^{(3)}_{i_4 i_5 i_6)}$,
$\delta_{(i_1 i_2} \delta_{i_4 i_5} \delta_{i_5 i_6)}$
(${t}^{(3)}_{i_1 i_2 i_3}$ is proportional to $d_{i_1 i_2 i_3}$,
see Sec.\ref{suse6.2}).
\end{example}
As a consequence of Lemma \ref{lem3.2}, it follows that we can obtain
Casimir operators ${\cal C}'$ from cocycles by means of

\begin{corollary}
({\it Generalised Casimirs from primitive cocycles})
\\
The operator
\be
{\cal C}'^{(m)}=
[\Omega^{(2m-1)}]^{j_1 \dots j_{2m-1}} X_{j_1} \dots X_{j_{2m-1}}
\label{seIIxiii}
\ee
is an $m$-order Casimir operator for $\g$.
\label{cor3.2}
\end{corollary}
{\it Proof}:\quad
Using the skewsymmetry of $\Omega^{(2m-1)}$ we rewrite (\ref{seIIxiii}) in the
form
\be
\begin{array}{l}
\displaystyle
{\cal C}'^{(m)}=
[\Omega^{(2m-1)}]^{j_1 \dots j_{2m-1}} X_{j_1} \dots X_{j_{2m-1}}
\\[0.3cm]
\qquad
\displaystyle
=
{1\over 2^{m-1}} [\Omega^{(2m-1)}]^{j_1 \dots j_{2m-2} \sigma}
[X_{j_1},X_{j_2}] \dots  [X_{j_{2m-3}} , X_{j_{2m-2}}]  X_{\sigma}
\\[0.3cm]
\qquad
\displaystyle
=
{1\over 2^{m-1}}
[\Omega^{(2m-1)}]^{j_1 \dots j_{2m-2} \sigma}
C^{i_1}_{j_1 j_2} \dots C^{i_{m-1}}_{j_{2m-3} j_{2m-2}}
X_{i_1} \dots X_{i_{m-1}} X_{\sigma}
\\[0.3cm]
\displaystyle
\qquad
={1\over 2^{m-1}}
{t}^{i_1 \dots i_{m-1} \sigma} X_{i_1}
\dots X_{i_{m-1}} X_{\sigma}
\end{array}
\label{nuevocas}
\ee
and it follows that it is a Casimir as a consequence of Lemma \ref{lem2.1},
{\it q.e.d.}

\section{The case of $su(n)$: coordinates of the $su(3)$- and $su(4)$-cocycles}
\label{cocycles}

  Let us take as a basis $\{T_i\}$ of the $su(n)$ algebra the $(n^2-1)$
traceless and  hermitian $n\times n$ matrices of the defining representation
of $su(n)$ satisfying the relations
\be
\begin{array}{l}
\displaystyle
[T_i,T_j]=i{f_{ij}}_\cdot ^k T_k\quad,\quad
\{T_i,T_j\}= c\delta_{ij}+ {d_{ij}}_\cdot^k T_k\quad,
\\[0.3cm]
\displaystyle
\mbox{Tr} ( T_i T_j) ={1\over 2} \delta_{ij} \quad,\quad
T_i T_j ={c\over 2}\delta_{ij} + {1\over 2}{d_{ij}}_\cdot^k
T_k
+{i\over 2}{f_{ij}}_\cdot^k T_k\quad,
\end{array}
\label{seIIaai}
\ee
where $i,j,k=1,\dots,n^2-1$ and $c\equiv {1\over n}$ (for a study
of the $su(3)$ tensors, see \cite{MSW}). In the `physical' (Gell-Mann)
basis, it is customary to use the $\lambda$-matrices $\lambda_i=2T_i$
for which the above relations trivially become

\be
\begin{array}{l}
\displaystyle
[\lambda_i,\lambda_j]=2i{f_{ij}}_\cdot ^k \lambda_k\quad,\quad
\{\lambda_i,\lambda_j\}= 4c\delta_{ij}+ 2{d_{ij}}_\cdot^k\lambda_k\quad,
\\[0.3cm]
\displaystyle
\mbox{Tr} (\lambda_i \lambda_j) =2 \delta_{ij} \quad,\quad
\lambda_i\lambda_j =2c\delta_{ij} + ({d_{ij}}_\cdot^k +i {f_{ij}}_\cdot^k)
\lambda_k\quad.
\end{array}
\label{seIIai}
\ee
Using the Gell-Mann representation of $su(3)$ we have

\begin{table}
\caption{Non-zero structure constants for $su(3)$.}\label{table1}
$$
  \begin{array}{lll}
    f_{123} = 1 &
    f_{147} = 1/2 &
    f_{156} = -1/2 \\
    f_{246} = 1/2 &
    f_{257} = 1/2 &
    f_{345} = 1/2 \\
    f_{367} = -1/2 &
    f_{458} = \sqrt{3}/2 &
    f_{678} = \sqrt{3}/2
  \end{array}
$$
\end{table}
\begin{table}
\caption{$3rd$-order invariant symmetric polynomial for $su(3)$.}\label{table3}
$$
\begin{array}{lllll}
d_{118}=1/\sqrt{3} &
d_{228}=1/\sqrt{3} &
d_{338}=1/\sqrt{3} &
d_{888}=-1/\sqrt{3} &
\\
d_{448}=-1/(2\sqrt{3}) &
d_{558}=-1/(2\sqrt{3}) &
d_{668}=-1/(2\sqrt{3}) &
d_{778}=-1/(2\sqrt{3}) &
\\
d_{146}=1/2 &
d_{157}=1/2 &
d_{247}=-1/2 &
d_{256}=1/2 &
\\
d_{344}=1/2 &
d_{355}=1/2 &
d_{366}=-1/2 &
d_{377}=-1/2 &
\end{array}
$$
\end{table}
The $f_{ijk}$ constitute the coordinates of the $su(3)$ three-cocycle.
If we use the structure constants $f_{ijk}$ and the symmetric $d_{ijk}$
to define [{\it c.f.} (\ref{seIIviii})] the coordinates of the
five-cocycle $\Omega^{(5)}$ by

\begin{equation}
  \Omega^{(5)}_{i_1 i_2 i_3 i_4 i_5} =
{f^j}_{i_1 [i_2} {f^k}_{i_3 i_4]} d_{j k i_5} \quad,
\label{omega5}
\end{equation}
we obtain

\begin{table}
\caption{Non-zero coordinates of the $su(3)$ five-cocycle.}\label{table2}
$$
\begin{array}{lcllcllcl}
    \Omega_{12345} & = & 1/4 &
    \Omega_{12367} & = & 1/4 &
    \Omega_{12458} & = & \sqrt{3}/12, \\
    \Omega_{12678} & = & -\sqrt{3}/12 &
    \Omega_{13468} & = & -\sqrt{3}/12 &
    \Omega_{13578} & = & -\sqrt{3}/12 , \\
    \Omega_{23478} & = & \sqrt{3}/12 &
    \Omega_{23568} & = & -\sqrt{3}/12 &
    \Omega_{45678} & = & -\sqrt{3}/6 .
\end{array}
$$
\end{table}
Using the natural extension of the Gell-Mann labelling
of generators to $su(4)$ in agreement with the contents of Table VI in
\cite{HIIST} (where in its second part $f$ should be replaced by $d$),
we have

\begin{table}
\caption{Non-zero structure constants for $su(4)$.}\label{table4}
$$
  \begin{array}{lll}
    f_{1,2,3} = 1 & f_{1,4,7} = 1/2 & f_{1,5,6} = -1/2 \\
    f_{1,9,12} = 1/2 & f_{1,10,11} = -1/2 & f_{2,4,6} = 1/2 \\
    f_{2,5,7} = 1/2 & f_{2,9,11} = 1/2 & f_{2,10,12} = 1/2 \\
    f_{3,4,5} = 1/2 & f_{3,6,7} = -1/2 & f_{3,9,10} = 1/2 \\
    f_{3,11,12} = -1/2 & f_{4,5,8} = \sqrt{3}/2 & f_{4,9,14} = 1/2 \\
    f_{4,10,13} = -1/2 & f_{5,9,13} = 1/2 & f_{5,10,14} = 1/2 \\
    f_{6,7,8} = \sqrt{3}/2 & f_{6,11,14} = 1/2 & f_{6,12,13} = -1/2 \\
    f_{7,11,13} = 1/2 & f_{7,12,14} = 1/2 & f_{8,9,10} = 1/(2 \sqrt{3}) \\
    f_{8,11,12} = 1/(2 \sqrt{3}) & f_{8,13,14} = -1/\sqrt{3} &
      f_{9,10,15} = \sqrt{2}/\sqrt{3} \\
    f_{11,12,15} = \sqrt{2}/\sqrt{3} & f_{13,14,15} = \sqrt{2}/\sqrt{3} &
  \end{array}
$$
\end{table}

\begin{table}
\caption{$3rd$-order invariant symmetric polynomial for $su(4)$.}\label{table5}
$$
\begin{array}{llll}
d_{4,4,3} = 1/2 &
d_{5,5,3} = 1/2 &
d_{6,6,3} = -1/2 &
d_{7,7,3} = -1/2
\\
d_{9,9,3} = 1/2 &
d_{10,10,3} = 1/2 &
d_{11,11,3} = -1/2 &
d_{12,12,3} = -1/2
\\
d_{1,1,8} = 1/\sqrt{3} &
d_{2,2,8} = 1/\sqrt{3} &
d_{3,3,8} = 1/\sqrt{3} &
d_{4,4,8} = -1/(2\sqrt{3})
\\
d_{5,5,8} = -1/(2\sqrt{3}) &
d_{6,6,8} = -1/(2\sqrt{3}) &
d_{7,7,8} = -1/(2\sqrt{3}) &
d_{8,8,8} = -1/\sqrt{3}
\\
d_{9,9,8} = 1/(2\sqrt{3}) &
d_{10,10,8} = 1/(2\sqrt{3}) &
d_{11,11,8} = 1/(2\sqrt{3}) &
d_{12,12,8} = 1/(2\sqrt{3})
\\
d_{13,13,8} = -1/\sqrt{3} &
d_{14,14,8} = -1/\sqrt{3} &
d_{1,1,15} = 1/\sqrt{6} &
d_{2,2,15} = 1/\sqrt{6}
\\
d_{3,3,15} = 1/\sqrt{6} &
d_{4,4,15} = 1/\sqrt{6} &
d_{5,5,15} = 1/\sqrt{6} &
d_{6,6,15} = 1/\sqrt{6}
\end{array}
$$
\end{table}
$$
\begin{array}{llll}
d_{7,7,15} = 1/\sqrt{6} &
d_{8,8,15} = 1/\sqrt{6} &
d_{9,9,15} = -1/\sqrt{6} &
d_{10,10,15} = -1/\sqrt{6}
\\
d_{11,11,15} = -1/\sqrt{6} &
d_{12,12,15} = -1/\sqrt{6} &
d_{13,13,15} = -1/\sqrt{6} &
d_{14,14,15} = -1/\sqrt{6}
\\
d_{15,15,15} = -2/\sqrt{6} &
d_{1,4,6} = 1/2 &
d_{1,5,7} = 1/2 &
d_{1,9,11} = 1/2
\\
d_{1,10,12} = 1/2 &
d_{2,4,7} = -1/2 &
d_{2,5,6} = 1/2 &
d_{2,9,12} = -1/2
\\
d_{2,10,11} = 1/2 &
d_{4,9,13} = 1/2 &
d_{4,10,14} = 1/2 &
d_{5,9,14} = -1/2
\\
d_{5,10,13} = 1/2 &
d_{6,11,13} = 1/2 &
d_{6,12,14} = 1/2 &
d_{7,11,14} = -1/2
\\
d_{7,12,13} = 1/2 & & & \\
\end{array}
$$
\hrule

\begin{table}
\caption{Non-zero coordinates of the $su(4)$ five-cocycle.}\label{table6}
$$
\begin{array}{lll}
  \Omega_{1,2,3,4,5} = 1/4 &
    \Omega_{1,2,3,6,7} = 1/4 &
    \Omega_{1,2,3,9,10} = 1/4 \\
  \Omega_{1,2,3,11,12} = 1/4 &
    \Omega_{1,2,4,5,8} = \sqrt{3}/12 &
    \Omega_{1,2,4,9,14} = 1/12 \\
  \Omega_{1,2,4,10,13} = -1/12 &
    \Omega_{1,2,5,9,13} = 1/12 &
    \Omega_{1,2,5,10,14} = 1/12 \\
  \Omega_{1,2,6,7,8} = -\sqrt{3}/12 &
    \Omega_{1,2,6,11,14} = -1/12 &
    \Omega_{1,2,6,12,13} = 1/12 \\
  \Omega_{1,2,7,11,13} = -1/12 &
    \Omega_{1,2,7,12,14} = -1/12 &
    \Omega_{1,2,8,9,10} = \sqrt{3}/36 \\
  \Omega_{1,2,8,11,12} = -\sqrt{3}/36 &
    \Omega_{1,2,9,10,15} = \sqrt{6}/18 &
    \Omega_{1,2,11,12,15} = -\sqrt{6}/18 \\
  \Omega_{1,3,4,6,8} = -\sqrt{3}/12 &
    \Omega_{1,3,4,11,13} = 1/12 &
    \Omega_{1,3,4,12,14} = 1/12 \\
  \Omega_{1,3,5,7,8} = -\sqrt{3}/12 &
    \Omega_{1,3,5,11,14} = -1/12 &
    \Omega_{1,3,5,12,13} = 1/12 \\
  \Omega_{1,3,6,9,13} = -1/12 &
    \Omega_{1,3,6,10,14} = -1/12 &
    \Omega_{1,3,7,9,14} = 1/12 \\
  \Omega_{1,3,7,10,13} = -1/12 &
    \Omega_{1,3,8,9,11} = -\sqrt{3}/36 &
    \Omega_{1,3,8,10,12} = -\sqrt{3}/36
\end{array}
$$
\end{table}
$$ \begin{array}{lll}
  \Omega_{1,3,9,11,15} = -\sqrt{6}/18 &
    \Omega_{1,3,10,12,15} = -\sqrt{6}/18 &
    \Omega_{1,4,5,9,12} = 1/12 \\
  \Omega_{1,4,5,10,11} = -1/12 &
    \Omega_{1,4,7,9,10} = 1/12 &
    \Omega_{1,4,7,11,12} = 1/12 \\
  \Omega_{1,4,7,13,14} = 1/12 &
    \Omega_{1,4,8,11,13} = -\sqrt{3}/36 &
    \Omega_{1,4,8,12,14} = -\sqrt{3}/36 \\
  \Omega_{1,4,11,13,15} = \sqrt{6}/36 &
    \Omega_{1,4,12,14,15} = \sqrt{6}/36 &
    \Omega_{1,5,6,9,10} = -1/12 \\
  \Omega_{1,5,6,11,12} = -1/12 &
    \Omega_{1,5,6,13,14} = -1/12 &
    \Omega_{1,5,8,11,14} = \sqrt{3}/36 \\
  \Omega_{1,5,8,12,13} = -\sqrt{3}/36 &
    \Omega_{1,5,11,14,15} = -\sqrt{6}/36 &
    \Omega_{1,5,12,13,15} = \sqrt{6}/36 \\
  \Omega_{1,6,7,9,12} = 1/12 &
    \Omega_{1,6,7,10,11} = -1/12 &
    \Omega_{1,6,8,9,13} = -\sqrt{3}/36 \\
  \Omega_{1,6,8,10,14} = -\sqrt{3}/36 &
    \Omega_{1,6,9,13,15} = \sqrt{6}/36 &
    \Omega_{1,6,10,14,15} = \sqrt{6}/36 \\
  \Omega_{1,7,8,9,14} = \sqrt{3}/36 &
    \Omega_{1,7,8,10,13} = -\sqrt{3}/36 &
    \Omega_{1,7,9,14,15} = -\sqrt{6}/36 \\
  \Omega_{1,7,10,13,15} = \sqrt{6}/36 &
    \Omega_{1,9,12,13,14} = -1/12 &
    \Omega_{1,10,11,13,14} = 1/12 \\
\end{array}
$$
\hrule
$$ \begin{array}{lll}
  \Omega_{2,3,4,7,8} = \sqrt{3}/12 &
    \Omega_{2,3,4,11,14} = 1/12 &
    \Omega_{2,3,4,12,13} = -1/12 \\
  \Omega_{2,3,5,6,8} = -\sqrt{3}/12 &
    \Omega_{2,3,5,11,13} = 1/12 &
    \Omega_{2,3,5,12,14} = 1/12 \\
  \Omega_{2,3,6,9,14} = 1/12 &
    \Omega_{2,3,6,10,13} = -1/12 &
    \Omega_{2,3,7,9,13} = 1/12 \\
  \Omega_{2,3,7,10,14} = 1/12 &
    \Omega_{2,3,8,9,12} = \sqrt{3}/36 &
    \Omega_{2,3,8,10,11} = -\sqrt{3}/36 \\
  \Omega_{2,3,9,12,15} = \sqrt{6}/18 &
    \Omega_{2,3,10,11,15} = -\sqrt{6}/18 &
    \Omega_{2,4,5,9,11} = 1/12 \\
  \Omega_{2,4,5,10,12} = 1/12 &
    \Omega_{2,4,6,9,10} = 1/12 &
    \Omega_{2,4,6,11,12} = 1/12 \\
  \Omega_{2,4,6,13,14} = 1/12 &
    \Omega_{2,4,8,11,14} = -\sqrt{3}/36 &
    \Omega_{2,4,8,12,13} = \sqrt{3}/36 \\
  \Omega_{2,4,11,14,15} = \sqrt{6}/36 &
    \Omega_{2,4,12,13,15} = -\sqrt{6}/36 &
    \Omega_{2,5,7,9,10} = 1/12 \\
  \Omega_{2,5,7,11,12} = 1/12 &
    \Omega_{2,5,7,13,14} = 1/12 &
    \Omega_{2,5,8,11,13} = -\sqrt{3}/36 \\
  \Omega_{2,5,8,12,14} = -\sqrt{3}/36 &
    \Omega_{2,5,11,13,15} = \sqrt{6}/36 &
    \Omega_{2,5,12,14,15} = \sqrt{6}/36 \\
  \end{array}
$$
\hrule
$$ \begin{array}{lll}
  \Omega_{2,6,7,9,11} = 1/12 &
    \Omega_{2,6,7,10,12} = 1/12 &
    \Omega_{2,6,8,9,14} = \sqrt{3}/36 \\
  \Omega_{2,6,8,10,13} = -\sqrt{3}/36 &
    \Omega_{2,6,9,14,15} = -\sqrt{6}/36 &
    \Omega_{2,6,10,13,15} = \sqrt{6}/36 \\
  \Omega_{2,7,8,9,13} = \sqrt{3}/36 &
    \Omega_{2,7,8,10,14} = \sqrt{3}/36 &
    \Omega_{2,7,9,13,15} = -\sqrt{6}/36 \\
  \Omega_{2,7,10,14,15} = -\sqrt{6}/36 &
    \Omega_{2,9,11,13,14} = -1/12 &
    \Omega_{2,10,12,13,14} = -1/12 \\
  \Omega_{3,4,5,9,10} = 1/6 &
    \Omega_{3,4,5,13,14} = 1/12 &
    \Omega_{3,4,8,9,13} = -\sqrt{3}/36 \\
  \Omega_{3,4,8,10,14} = -\sqrt{3}/36 &
    \Omega_{3,4,9,13,15} = \sqrt{6}/36 &
    \Omega_{3,4,10,14,15} = \sqrt{6}/36 \\
  \Omega_{3,5,8,9,14} = \sqrt{3}/36 &
    \Omega_{3,5,8,10,13} = -\sqrt{3}/36 &
    \Omega_{3,5,9,14,15} = -\sqrt{6}/36 \\
  \Omega_{3,5,10,13,15} = \sqrt{6}/36 &
    \Omega_{3,6,7,11,12} = -1/6 &
    \Omega_{3,6,7,13,14} = -1/12 \\
  \Omega_{3,6,8,11,13} = \sqrt{3}/36 &
    \Omega_{3,6,8,12,14} = \sqrt{3}/36 &
    \Omega_{3,6,11,13,15} = -\sqrt{6}/36 \\
  \Omega_{3,6,12,14,15} = -\sqrt{6}/36 &
    \Omega_{3,7,8,11,14} = -\sqrt{3}/36 &
    \Omega_{3,7,8,12,13} = \sqrt{3}/36 \\
\end{array} $$
\hrule
$$ \begin{array}{lll}
  \Omega_{3,7,11,14,15} = \sqrt{6}/36 &
    \Omega_{3,7,12,13,15} = -\sqrt{6}/36 &
    \Omega_{3,9,10,13,14} = -1/12 \\
  \Omega_{3,11,12,13,14} = 1/12 &
    \Omega_{4,5,6,7,8} = -\sqrt{3}/6 &
    \Omega_{4,5,6,11,14} = -1/12 \\
  \Omega_{4,5,6,12,13} = 1/12 &
    \Omega_{4,5,7,11,13} = -1/12 &
    \Omega_{4,5,7,12,14} = -1/12 \\
  \Omega_{4,5,8,9,10} = \sqrt{3}/9 &
    \Omega_{4,5,8,13,14} = 5 \sqrt{3}/36 &
    \Omega_{4,5,9,10,15} = \sqrt{6}/18 \\
  \Omega_{4,5,13,14,15} = -\sqrt{6}/18 &
    \Omega_{4,6,7,9,14} = -1/12 &
    \Omega_{4,6,7,10,13} = 1/12 \\
  \Omega_{4,6,8,9,11} = \sqrt{3}/18 &
    \Omega_{4,6,8,10,12} = \sqrt{3}/18 &
    \Omega_{4,6,9,11,15} = \sqrt{6}/36 \\
  \Omega_{4,6,10,12,15} = \sqrt{6}/36 &
    \Omega_{4,7,8,9,12} = \sqrt{3}/18 &
    \Omega_{4,7,8,10,11} = -\sqrt{3}/18 \\
  \Omega_{4,7,9,12,15} = \sqrt{6}/36 &
    \Omega_{4,7,10,11,15} = -\sqrt{6}/36 &
    \Omega_{4,8,9,13,15} = -\sqrt{2}/12 \\
  \Omega_{4,8,10,14,15} = -\sqrt{2}/12 &
    \Omega_{4,9,11,12,14} = -1/12 &
    \Omega_{4,10,11,12,13} = 1/12 \\
  \Omega_{5,6,7,9,13} = -1/12 &
    \Omega_{5,6,7,10,14} = -1/12 &
    \Omega_{5,6,8,9,12} = -\sqrt{3}/18 \\
\end{array} $$
\hrule
$$ \begin{array}{lll}
  \Omega_{5,6,8,10,11} = \sqrt{3}/18 &
    \Omega_{5,6,9,12,15} = -\sqrt{6}/36 &
    \Omega_{5,6,10,11,15} = \sqrt{6}/36 \\
  \Omega_{5,7,8,9,11} = \sqrt{3}/18 &
    \Omega_{5,7,8,10,12} = \sqrt{3}/18 &
    \Omega_{5,7,9,11,15} = \sqrt{6}/36 \\
  \Omega_{5,7,10,12,15} = \sqrt{6}/36 &
    \Omega_{5,8,9,14,15} = \sqrt{2}/12 &
    \Omega_{5,8,10,13,15} = -\sqrt{2}/12 \\
  \Omega_{5,9,11,12,13} = -1/12 &
    \Omega_{5,10,11,12,14} = -1/12 &
    \Omega_{6,7,8,11,12} = \sqrt{3}/9 \\
  \Omega_{6,7,8,13,14} = 5 \sqrt{3}/36 &
    \Omega_{6,7,11,12,15} = \sqrt{6}/18 &
    \Omega_{6,7,13,14,15} = -\sqrt{6}/18 \\
  \Omega_{6,8,11,13,15} = -\sqrt{2}/12 &
    \Omega_{6,8,12,14,15} = -\sqrt{2}/12 &
    \Omega_{6,9,10,11,14} = -1/12 \\
  \Omega_{6,9,10,12,13} = 1/12 &
    \Omega_{7,8,11,14,15} = \sqrt{2}/12 &
    \Omega_{7,8,12,13,15} = -\sqrt{2}/12 \\
  \Omega_{7,9,10,11,13} = -1/12 &
    \Omega_{7,9,10,12,14} = -1/12 &
    \Omega_{8,9,10,11,12} = -\sqrt{3}/18 \\
  \Omega_{8,9,10,13,14} = \sqrt{3}/36 &
    \Omega_{8,11,12,13,14} = \sqrt{3}/36 &
    \Omega_{9,10,11,12,15} = -\sqrt{6}/9 \\
  \Omega_{9,10,13,14,15} = -\sqrt{6}/9 &
    \Omega_{11,12,13,14,15} = -\sqrt{6}/9 & \\
\end{array}
$$
\hrule

\medskip
Any pair of sets of non-zero coordinates ($i_1,i_2,i_3$)
and ($i_4,i_5,i_6,i_7,i_8$)
for the $su(4)$ three- and five-cocycles with distinct index sets
defines a non-zero coordinate of the seven-cocycle, the indices
($i_9,\dots,i_{15}$) of which take the remaining available values
in the set $i$=$(1,\cdots,15)$ [see (\ref{ochodos})
below]. There are more than 400 non-zero such coordinates,
which are not given here.

\section{Primitive invariant symmetric polynomials, \\ Casimirs and cocycles}
\label{se4}

\subsection{General considerations and the case of $su(n)$}
\label{suse51}

The polynomial ring of commuting operators in the enveloping algebra
${\cal U}(\g)$ of a simple algebra $\g$ is freely generated by $l$
Casimir-Racah operators \cite{RACAH,GEL,LCB,AK,GR,PP,OKPA,SOK} of
orders $m_1,\dots,m_l$. As a result, the $k^{(m)}$ polynomials for
$su(n)$, say, will be expressible in terms of the lower order
primitive ones if $m>n=l+1$.  For instance, $k_{i_1 i_2 i_3 i_4}=
{1\over 4!}\mbox{sTr} (X_{i_1} X_{i_2} X_{i_3} X_{i_4})$ for $su(4)$
is primitive and generates a (non-trivial) fourth-order Casimir, but
it turns out to be proportional to $\delta_{(i_1 i_2} \delta_{i_3
  i_4)}$ (and does not lead to a fourth-order primitive Casimir) when
the $T_i\in su(3)$ (see Example \ref{ex3.1} below and
(\ref{seVIvii})).  The case of $su(l+1)$ has been considered before
(see, \eg, \cite{RS,BBSS,SUDBERY}) but to our knowledge there is no
unified treatment available in the literature for the four simple
infinite series.

Let us start by recalling the case of $su(n)$ \cite{RS} and consider
\be
\epsilon_{\alpha_1 \dots \alpha_{n+1}}^{\beta_1\dots\beta_{n+1}}
(T_{i_1})^{\alpha_1}_{\cdot\beta_1} \cdots
(T_{i_{n+1}})^{\alpha_{n+1}}_{\cdot\beta_{n+1}}=0\quad.
\label{seIIaii}
\ee
This expression is symmetric in the generator indices $i_1\dots i_{n+1}$ and
is obviously zero since the matrix representation indices
$\alpha,\beta$ range from 1 to $n$. Using (\ref{seIIaiii}),
it follows that the above expression is a sum of $(n+1)!$-terms which may be
grouped in classes, each class being a
sum of terms all involving a product of the same number
$\nu_1,\dots,\nu_{n+1}$ of products
of traces of products of $1,\dots,n+1$
matrices respectively,
where $1\cdot\nu_1+2\cdot\nu_2+\dots+(n+1)\nu_{n+1}=(n+1)$.
In other words, the different {\it types} of products appearing
in (\ref{seIIaiii}) are characterised by the partitions
of $(n+1)$ elements \ie\ by the Young patterns (see, \eg, \cite{Hamermesh})
associated with $S_{n+1}$,
\be
[(n+1)^{\nu_{n+1}},n^{\nu_n},\dots,2^{\nu_2},1^{\nu_1}]
\label{seIIaiiia}
\ee
(obviously, $\nu_{n+1}=1$ or 0).
All the elements of $S_{n+1}$ for a given pattern (a set of fixed integers
$\nu_1,\dots,\nu_{n+1}$) determine products of
traces of matrices with the same grouping pattern and, moreover, appear in
(\ref{seIIaii}) with the same sign (they correspond in $S_{n+1}$ to the same
conjugation class).
The number of $S_N$ elements associated with
a given Young pattern is given by the 1844 Cauchy formula

\be
{ N! \over \nu_1 ! 2^{\nu_2} \nu_2 ! 3^{\nu_3} \nu_3 ! \dots N^{\nu_N} }
\label{seIIaiv}
\ee
and since they correspond to permutations $\sigma\in S_N$ with (equal) parity
$\pi(\sigma)$,
\be
\pi(\sigma)=(-1)^{\nu_2 + \nu_4 + \nu_6 + \dots}
\quad,
\label{seIIav}
\ee
they all contribute to (\ref{seIIaii}) with the same sign.

The mechanism which expresses the higher-order symmetric invariant polynomials
in terms of the primitive ones is now clear.
A given Young pattern determines a specific product of invariant symmetric
tensors with the sign (\ref{seIIav}) and weighted by the factor
(\ref{seIIaiv}).
Since one of the terms in the sum (\ref{seIIaii}) corresponds to the partition
$\nu_1=\dots=\nu_{N-1}=0\,,\,\nu_N=1$, it follows that the invariant symmetric
tensor of order $m=n+1=l+2$ will be expressed, through (\ref{seIIaii}),
in terms of the $l$ tensors of order $2,3,\dots,l+1$ and that only these
are primitive for $n=l+1$.

We are thus lead to the following

\begin{lemma}
({\it Invariant symmetric polynomials on $su(n)$})
\\
There are $l$ invariant polynomials of order $2,3,\dots,l+1$; the others are
not primitive, and may be expressed in terms of products of them.
\end{lemma}

The simplest application is that $d_{i_1 i_2 i_3}=0$ for $su(2)$.
For the next higher order we have the following

\begin{example}\
\label{ex3.1}
\\
Let $\g=su(3)$. Using the $\lambda_i$ matrices, ($i=1,\dots,8$), we find,
since they are tridimensional,
$$
\begin{array}{l}
\displaystyle
\epsilon^{\beta_1 \beta_2 \beta_3 \beta_4}_{\alpha_1 \alpha_2 \alpha_3
\alpha_4}
(\lambda_{i_1})^{\alpha_1}_{\cdot\beta_1}
(\lambda_{i_2})^{\alpha_2}_{\cdot\beta_2}
(\lambda_{i_3})^{\alpha_3}_{\cdot\beta_3}
(\lambda_{i_4})^{\alpha_4}_{\cdot\beta_4}
\\[0.3cm]
\displaystyle
=
\mbox{s}\{{1\over 2^2 2}\mbox{Tr}(\lambda_{i_1}\lambda_{i_2})
\mbox{Tr}(\lambda_{i_3}\lambda_{i_4})
-{1\over 4}
\mbox{Tr}(\lambda_{i_1}\lambda_{i_2}\lambda_{i_3}\lambda_{i_4}) \} =0 \quad,
\end{array}
$$
where, as before, $\mbox{s}$ means symmetrisation in all indices
$i_1,i_2,i_3,i_4$. Thus, the fourth-order symmetric tensor can be
expressed in terms of the Killing second-order one. Using
(\ref{seIIai}) we find

\be
\mbox{Tr}(\lambda_{(i_1}\lambda_{i_2}\lambda_{i_3}\lambda_{i_4)})
= 2 \delta_{(i_1 i_2} \delta_{i_3 i_4)}
\quad,
\label{seIIaa}
\ee
But using again (\ref{seIIai}) we may compute directly for $su(n)$

\be
{1\over 4!}\mbox{sTr}(
\lambda_{i_1}\lambda_{i_2}\lambda_{i_3}\lambda_{i_4})=
{1\over 4\cdot 4!}\mbox{sTr}(\{\lambda_{i_1},\lambda_{i_2}\}
\{\lambda_{i_3},\lambda_{i_4}\})=
{4\over n} \delta_{(i_1 i_2} \delta_{i_3 i_4)} + 2 d_{\rho (i_1 i_2}
d_{i_3 i_4) \rho}
\ .
\label{seIIab}
\ee
Eq. (\ref{seIIaa}) and eq. (\ref{seIIab}) for $n=3$ now give
$d_{\rho (i_1 i_2} d_{i_3 i_4) \rho} = {1\over 3}
\delta_{(i_1 i_2} \delta_{i_3 i_4)}.$ For $su(3)$, the tensor
$k_{i_1 i_2 i_3 i_ 4}$ [eq. (\ref{seIi})] is given by

\be
k_{i_1 i_2 i_3 i_4}= \left( {-i\over 2} \right)^4
\mbox{Tr}(\lambda_{(i_1}\lambda_{i_2}\lambda_{i_3}\lambda_{i_4)})
={1\over 12}\delta_{(i_1 i_2}\delta_{i_3 i_4)}
+ {1\over 8} d_{\rho (i_1 i_2}d_{i_3 i_4)\rho} \quad .
\label{newsu3ten}
\ee

\end{example}

\subsection{The case of $so(n)$ and $sp(l)$}
\label{suse52}

Let us now turn to the case of the orthogonal (odd, even) $(B_l,D_l)$
and symplectic $(C_l)$ algebras. In the defining representation these
groups preserve the $n\times n$ euclidean or symplectic metric $\eta$
and thus the generators $X_i$ of these algebras satisfy
\be
X_i \eta=-\eta X_i^t\quad,
\label{seIIavi}
\ee
where $i=1,\dots,l(2l+1)$, $l\ge 2$, $n=2l+1$ ($B_l$);
$i=1,\dots,l(2l+1)$, $l\ge 3$, $n=2l$ ($C_l$);
and $i=1,\dots,l(2l-1)$, $l\ge 4$, $n=2l$ ($D_l$).

The symmetrised products of an {\it odd} number of generators of the
orthogonal and symplectic groups also satisfies (\ref{seIIavi});
hence, they are a member of their respective algebras. In particular,
we may write

\be
\{X_{i_1},X_{i_2},X_{i_3}\}={v_{i_1 i_2 i_3}}^\sigma_\cdot X_\sigma
\quad,
\label{seIIavii}
\ee
where the bracket $\{\ ,\dots,\ \}$ denotes symmetrisation \ie, it is the sum
of the $6$ possible products and $v_{i_1 i_2 i_3 \sigma}$ is an
invariant symmetric polynomial \cite{HIGHER}.
Now we may define a new $v^{(m)}$ family of invariant symmetric polynomials
(cf. $d^{(m)}$ used in (\ref{VIai}) for $su(n)$) by

\be
v_{i_1\dots i_{2p}}=
{v_{(i_1 i_2 i_3}}^{\alpha_1}_\cdot
{v_{\alpha_1 i_4 i_5}}^{\alpha_2}_\cdot
\cdots {v_{\alpha_{p-2} i_{2p-2} i_{2p-1} i_{2p})}}
\label{hfamily}
\ee
With this notation (see (\ref{basickilling}))
\be
\begin{array}{rl}
k_{i_1\dots i_{2p}} &
\displaystyle
= {1\over (2p)!}\mbox{sTr} (X_{i_1}\cdots X_{i_{2p}}) \equiv
\mbox{Tr} (X_{(i_1}\cdots X_{i_{2p)}})
\\[0.3cm]
& =
\displaystyle
{1\over 6^{p-1}}
\mbox{Tr}(
\{\dots\{\{X_{(i_1},X_{i_2},X_{i_3}\},X_{i_4},X_{i_5}\},\dots,
X_{i_{2p-2}},X_{i_{2p-1}}\}X_{i_{2p)}})
\\[0.3cm]
\displaystyle
&
\displaystyle
= {1\over 6^{p-1}}
{v_{(i_1 i_2 i_3}}^{\alpha_1}_\cdot
{v_{\alpha_1 i_4 i_5}}^{\alpha_2}_\cdot
\cdots {v_{\alpha_{p-2} i_{2p-2} i_{2p-1}}}^\sigma_\cdot
\mbox{Tr}(X_\sigma X_{i_{2p})})
\\[0.3cm]
\displaystyle
&
\displaystyle
= {\kappa\over 6^{p-1}}
{v_{(i_1 i_2 i_3}}^{\alpha_1}_\cdot
{v_{\alpha_1 i_4 i_5}}^{\alpha_2}_\cdot
\cdots {v_{\alpha_{p-2} i_{2p-2} i_{2p-1} i_{2p})}}
\quad.
\end{array}
\label{seIIaviia}
\ee
Thus,
\be
k_{i_1\dots i_{2p}} =
{\kappa \over 6^{p-1}} v_{i_1\dots i_{2p}}
\quad .
\label{seIIaviii}
\ee
This leads us to the following simple lemma

\begin{lemma}
({\it Generation of higher order invariant symmetric polynomials for
$B_l,C_l,D_l$})
\\
Let $v_{i_1 i_2 i_3 i_4}$ be the second lowest
symmetric invariant polynomial for $B_l,C_l,D_l$.
Then, the higher order symmetric polynomials $k_{i_1\dots i_{2p}}$
may be written
in terms of $v_{i_1 i_2 i_3 i_4}$ by means of (\ref{seIIaviii})
and (\ref{hfamily}).
\end{lemma}

Let now $\g$ be $B_l$ $(C_l)$ and let $X_i$ be a basis for $B_l$
($C_l$) given by $(2l+1)$- $(2l)$- dimensional matrices.  Then, since
the symmetric trace of a product of an odd number of $X$'s is zero,
and any partition of an odd number of elements will always include a
symmetric trace of an odd number of $X$'s we have to consider an even
product
\be
\epsilon_{\alpha_1 \dots \alpha_{2l+2}}^{\beta_1\dots\beta_{2l+2}}
(X_{i_1})^{\alpha_1}_{\cdot\beta_1}
(X_{i_2})^{\alpha_2}_{\cdot\beta_2} \cdots
(X_{i_{2l+2}})^{\alpha_{2l+2}}_{\cdot\beta_{2l+2}}=0\quad;
\label{seIIaix}
\ee
otherwise each term in the $l.h.s.$ would be zero.
Reasoning as before, but taking now into account that only traces of an {\it
even} number of factors are different from zero, we are led to the following

\begin{lemma}
\ ({\it Invariant symmetric polynomials for $B_l,C_l$})
\\
The symmetric invariant polynomials for $B_l,C_l$ given by
(\ref{seIi}) are all of even order $m=2k$ and non-primitive for $m > 2l$.
The relation which expresses the lowest non-primitive symmetric polynomial in
terms of the primitive ones follows from the equality
\be
\begin{array}{l}
\displaystyle
\sum_{\mbox{partitions}}
{(-1)^{\nu_2+\nu_4+\dots+\nu_{2l+2}}\over
2^{\nu_2} \nu_2! 4^{\nu_4} \nu_4! \dots (2l+2)^{\nu_{2l+2}}}
\\[0.3cm]
\displaystyle
\qquad\qquad\qquad
\cdot
\mbox{s} \left \lbrace
[\mbox{Tr}(X_{i_1} X_{i_2})]^{\nu_2}
[\mbox{Tr}
(X_{i_{2\nu_2+1}} X_{i_{2\nu_2+2}} X_{i_{2\nu_2+3}} X_{i_{2\nu_2+4}})]^{\nu_4}
\dots
\right \rbrace=0
\quad,
\end{array}
\label{seIIax}
\ee
where $\sum$ is extended over all partitions of $(2l+2)$ in (even) factors
(all Young patterns of $S_{2l+2}$) and
there is symmetrisation over all indices $i_1,\dots,i_{2l+2}$.
\end{lemma}

\begin{example}
For $B_2$ we obtain that
$$
\begin{array}{rl}
\displaystyle
\epsilon^{\beta_1 \dots \beta_6}_{\alpha_1\dots \alpha_6}
(X_{i_1})^{\alpha_1}_{\cdot \beta_1}\dots (X_{i_6})^{\alpha_6}_{\cdot \beta_6}
=
&
\displaystyle
\mbox{s}\{
-{1\over 3!2^3}\mbox{Tr}(X_{i_1}X_{i_2})\mbox{Tr}(X_{i_3}X_{i_4})
\mbox{Tr}(X_{i_5}X_{i_6})
\\[0.3cm]
+
&\displaystyle
{1\over 4\cdot 2}
\mbox{Tr}(X_{i_1}X_{i_2}X_{i_3}X_{i_4})\mbox{Tr}(X_{i_5}X_{i_6})
-{1\over 6} \mbox{Tr}(X_{i_1}\dots X_{i_6}) \}
\\[0.3cm]
=
& 0 \quad,
\end{array}
$$
relation which expresses the invariant polynomial of order 6,
$\mbox{Tr}(X_{i_1}\dots X_{i_6})$, in terms of those of order 2 and 4.
\end{example}

\begin{example}
For $C_3$ we have

$$
\begin{array}{l}
\displaystyle
\epsilon^{\beta_1 \dots \beta_8}_{\alpha_1\dots \alpha_8}
(X_{i_1})^{\alpha_1}_{\cdot \beta_1}\dots (X_{i_8})^{\alpha_8}_{\cdot \beta_8}
=
\mbox{s}
\left\lbrace {1\over 4! 2^4}
\mbox{Tr}(X_{i_1}X_{i_2})\mbox{Tr}(X_{i_3}X_{i_4})\mbox{Tr}(X_{i_5}X_{i_6})
\mbox{Tr}(X_{i_7}X_{i_8}) \right.
\\[0.3cm]
\displaystyle
-{1\over 2! 2^2 4} \mbox{Tr}(X_{i_1}X_{i_2})\mbox{Tr}(X_{i_3}X_{i_4})
\mbox{Tr}(X_{i_5}X_{i_6}X_{i_7}X_{i_8})
+{1\over 12}
\mbox{Tr}(X_{i_1}X_{i_2})\mbox{Tr}(X_{i_3}X_{i_4}X_{i_5}X_{i_6}X_{i_7}X_{i_8})
\\[0.3cm]
\displaystyle
\left.
+{1\over 2! 4^2} \mbox{Tr}(X_{i_1}X_{i_2}X_{i_3}X_{i_4})
\mbox{Tr}(X_{i_5}X_{i_6}X_{i_7}X_{i_8})
-{1\over 8}\mbox{Tr}(X_{i_1}X_{i_2}X_{i_3}X_{i_4}X_{i_5}X_{i_6}X_{i_7}X_{i_8})
\right\rbrace
=0\quad.
\end{array}
$$
This expression relates the eighth-order invariant symmetric polynomial with
those of lower degree.
\end{example}

The case of the even orthogonal $D_l$ is different because, in this
case, there is an invariant polynomial which is related to the square
root of the determinant of an even dimensional matrix
(the Pfaffian).

\begin{lemma}
({\it Invariant symmetric polynomials for $D_l$})
\\
The symmetric invariant polynomials for $D_l$ given by (\ref{seIi}) are of
even order and primitive for $m=2,4\dots,2l-2$.
The higher order polynomials are written in terms of those and the polynomial
Pf$_{i_1\dots i_l}$ constructed from the Pfaffian using that

\be
\epsilon_{\alpha_1 \dots \alpha_{2l}}^{\beta_1 \dots \beta_{2l}}
(X_{i_1})^{\alpha_1}_{\cdot\beta_1}\dots
(X_{i_{2l}})^{\alpha_{2l}}_{\cdot\beta_{2l}}
=\mbox{Pf}_{(i_1\dots i_l} \mbox{Pf}_{i_{l+1}\dots i_{2l})}
\ee
which follows from the fact that
$\mbox{det}(\lambda^i X_i) = (\mbox{Pf}(\lambda^i X_i))^2$ for the
$\displaystyle \left({2l \atop l}\right)$
skewsymmetric matrices $X_i$, and that these matrices constitute a
basis in the vector space of $2l\times 2l$ skewsymmetric matrices.
\end{lemma}

\begin{example}
For $D_4$ we have
$$
\begin{array}{l}
\displaystyle
\epsilon^{\beta_1 \dots \beta_8}_{\alpha_1\dots \alpha_8}
(X_{i_1})^{\alpha_1}_{\cdot \beta_1}\dots (X_{i_8})^{\alpha_8}_{\cdot \beta_8}=
\displaystyle
\mbox{s}
\left\lbrace {1\over 4! 2^4}
\mbox{Tr}(X_{i_1}X_{i_2})\mbox{Tr}(X_{i_3}X_{i_4})\mbox{Tr}(X_{i_5}X_{i_6})
\mbox{Tr}(X_{i_7}X_{i_8})
\right.
\\
\displaystyle
-{1\over 2! 2^2 4} \mbox{Tr}(X_{i_1}X_{i_2})\mbox{Tr}(X_{i_3}X_{i_4})
\mbox{Tr}(X_{i_5}X_{i_6}X_{i_7}X_{i_8})
+{1\over 12}
\mbox{Tr}(X_{i_1}X_{i_2})\mbox{Tr}(X_{i_3}X_{i_4}X_{i_5}X_{i_6}X_{i_7}X_{i_8})
\\
\displaystyle
\left.
+{1\over 2! 4^2} \mbox{Tr}(X_{i_1}X_{i_2}X_{i_3}X_{i_4})
\mbox{Tr}(X_{i_5}X_{i_6}X_{i_7}X_{i_8})
-{1\over 8}\mbox{Tr}(X_{i_1}X_{i_2}X_{i_3}X_{i_4}X_{i_5}X_{i_6}X_{i_7}X_{i_8})
\right\rbrace
\\[0.3cm]
=\mbox{Pf}_{(i_1\dots i_4} \mbox{Pf}_{i_{5}\dots i_{8})}
\end{array}
$$
\end{example}

\section{Invariant symmetric tensors for $su(n)$: a detailed study}
\label{invsym}

The case of $su(n)$ is specially important since $SU(n)$-invariant
tensors appear in many physical theories as {\it e.g.}, QCD.
The properties of the $su(n)$-algebra tensors have already been discussed
in \cite{KR,MSW,RS,BBSS,SUDBERY} (and tables for the $f_{ijk}$ and
the $d_{ijk}$ for $su(n)$ up to $n$=6 have been given in \cite{HIIST}),
but we need to perform here a more complete and systematic
study. The case of the cocycles was already discussed in Sec.\ref{cocycles}.
We consider now the symmetric tensors, exhibiting in particular how
the relations defining them for general $n$ produce primitive tensors up to
a given order $m_l$ and non-primitive ones when this order is exceeded.
To this aim, we shall use Lemma \ref{lem3.3} and its Corollary \ref{corbasis}.

\subsection{$d$-tensors}

\label{suse6.1}

We begin our study of totally symmetric tensors of arbitrary ranks for $su(n)$
by considering the $d$-tensor family \cite{SUDBERY} (see also \cite{AK}).
For ranks $r=2$ and 3 we have $d_{ij}^{(2)}=\delta_{ij}$,
$d_{ijk}^{(3)}=d_{ijk}$, where the latter is the well-known totally
symmetric and traceless $su(n)$ tensor, which exists for $n\ge 3$.
Higher order tensors are defined recursively via
\be
\begin{array}{c}
\displaystyle
d^{(r+1)}_{i_1\dots i_{r-1} i_r i_{r+1}} = {d^{(r)}_{i_1\dots i_{r-1}}}^j
d^{(3)}_{j i_r i_{r+1}}\quad,\quad r=3,4,\dots\quad,
\\[0.3cm]
\displaystyle
d_{i_1 \dots i_m}^{(m)}= d_{i_1 i_2}^{l_1} d_{l_1 i_3}^{l_2}
\dots d_{l_{m-3} i_{m-1} i_m}
\end{array}
\label{VIai}
\ee
$i=1,...,(n^2-1)$. For $r\ge 3$, the above tensors are not symmetric in all
their indices, so the required totally symmetric tensors are defined from
their symmetrisation which gives
\be
d^{(r)}_{(i_1\dots i_r)} \quad.
\label{seVIaii}
\ee
Due to the fact that $d_{ijk}$ is already symmetric, (\ref{seVIaii})
normally represents the sum of $p<r!$ distinct terms divided by $p$.
The construction defines a family of symmetric invariant tensors of order
$r$.

 We want to know the dimension and a basis for the vector space
${\cal V}^{(m)}$ of invariant symmetric tensors of a given order $m$
(Corollary \ref{corbasis}). For example,
for $n$ large enough dim${\cal V}^{(4)}=2$ and a
basis is provided by
\be
d^{(4)}_{(i_1 i_2 i_3 i_4)} = d^{(4)}_{(i_1 i_2 i_3) i_4}
\quad,\quad
\delta_{(i_1 i_2} \delta_{i_3 i_4)} = \delta_{(i_1 i_2} \delta_{i_3) i_4}
\quad.
\label{seVIi}\ee
Similarly, also for $n$ large enough
dim${\cal V}^{(5)}=2$ and a basis is provided by
\be
d^{(5)}_{(i_1 i_2 i_3 i_4 i_5)}
\quad,\quad
d_{(i_1 i_2 i_3} \delta_{i_4 i_5)}
\quad.
\label{seVIii}
\ee
In fact, dim${\cal V}^{(4)}=2$, only for $n\ge 4$, since if (and only if)
$n=3$, we have from Ex.\ref{ex3.1}
\be
d^{(4)}_{(i_1 i_2 i_3 i_4)} = {1\over 3} \delta_{(i_1 i_2} \delta_{i_3 i_4)}
\quad.
\label{seVIiii}
\ee

Similarly, dim${\cal V}^{(5)}=2$ only for $n\ge 5$, a basis being provided
by $d^{(5)}_{(i_1 i_2 i_3 i_4 i_5)}$ and $d_{(i_1 i_2 i_3}\delta_{i_4 i_5)}$.
For $n=4$, however,

\be
d^{(5)}_{(i_1 i_2 i_3 i_4 i_5)} = {2\over 3} d_{(i_1 i_2 i_3} \delta_{i_4 i_5)}
\quad,
\label{seVIiv}
\ee
while, for $n=3$, we have as a direct consequence of (\ref{seVIiii})
\be
d^{(5)}_{(i_1 i_2 i_3 i_4 i_5)} = {1\over 3} d_{(i_1 i_2 i_3} \delta_{i_4 i_5)}
\quad.
\label{seVIv}
\ee
These results will serve as a check on some
calculations done later. They are a consequence of Corollary
\ref{corbasis}; eqs. (\ref{seVIiii}) and (\ref{seVIiv}) also
follow from work in \cite{SUDBERY} and in \cite{RS}, both of which supply
values for dim${\cal V}^{(4)}$ and suggest possible bases to use for increasing
$r$.
The recursive procedure of \cite{SUDBERY}
actually supplies more information, but the work in \cite{RS} is also useful,
providing direct access to results for
each fixed $n$.

One way to see that not all families of symmetric invariant
tensors of a given order are equally useful arises when one uses
them to construct Casimir
operators by (\ref{seIiv})

\be
{\cal C}^{(r)}= d^{(r)}_{(i_1 \dots i_r)} X^{i_1} \dots X^{i_r}
\label{seVIvi}
\ee
where the $X_i$ are generators of $su(n)$.  In principle this
yields an arbitrary number of Casimir operators, one of each order $r
\geq 2$. But since $su(n)$ has $(n-1)$ primitive Casimir
operators of order $2,3,\dots,n$, it must be possible to express those
with order $r>n$ in terms of the primitive ones.  It is well known how
to do this for $su(3)$, where (\ref{seVIiii}), (\ref{seVIv}) imply

\be
\begin{array}{c}
\displaystyle
{\cal C}^{(4)}={1\over 3}\delta_{(i_1 i_2}\delta_{i_3 i_4)}
X^{i_1} X^{i_2} X^{i_3} X^{i_4}
={1\over 3} ({\cal C}^{(2)})^2 +
{1\over 9} f_{i_1 i_2 i_3} X^{i_1} X^{i_2} X^{i_3}
={1\over 3}({\cal C}^{(2)})^2 + {1\over 6} {\cal C}^{(2)}
\quad,
\\[0.3cm]
\displaystyle
{\cal C}^{(5)}={1\over 3} {\cal C}^{(2)} {\cal C}^{(3)} +
{1\over 4} {\cal C}^{(3)}\quad .
\end{array}
\label{seVIvii}
\ee
Similarly, for $su(4)$ eq. (\ref{seVIiv}) gives

\be
{\cal C}^{(5)}={2\over 3} {\cal C}^{(2)} {\cal C}^{(3)} + {2\over 3}
{\cal C}^{(3)}
\quad.
\label{seVIviii}
\ee It is not possible in practice to treat such matters explicitly or
even systematically for arbitrarily large $n,r$.  It is thus
convenient to replace the family of symmetrised $d$-tensors by a
family for which no similar difficulties arise.  This is achieved by
using the $t$-tensors introduced in Lemma \ref{lem3.2} and taking
advantage of their property (\ref{seIIxiia}).

\subsection{$t$-tensors}
\label{suse6.2}

Working within the $t$ family, we can build from
$t^{(m)}$ only one non-vanishing scalar quantity,
\be
K^{(m)}(n)=t^{(m)\, i_1\dots i_m}
t^{(m)}_{i_1\dots i_m}
\quad;
\label{seVIxxviii}
\ee
all other full contractions are zero by Lemma \ref{lem3.3}.
Since the standard $su(n)$ Gell-Mann matrices $\lambda_i$ allow easy
computation of the $d$-tensors, and a well-developed
technology to operate with them exists (which can be completed
when necessary with additional relations, see Appendix), we give the
expressions of our $t$-tensors in terms of the $d$-tensors in Sec.
\ref{suse6.1}. {}From (\ref{seIIvii}) we trivially obtain
\be
t_{i_1 i_2}= n\delta_{i_1 i_2} \quad .
\label{seVIxxix}
\ee
Similarly, for $m$=3,4 eqs. (\ref{seIIxii}), (\ref{seIIv}) give
\be
t_{i_1 i_2 i_3}={n^2\over 3} k_{i_1 i_2 i_3}=
                {i n^2\over 12} d_{ijk} \quad,
\label{seVIxxx}
\ee
\be
t_{i_1 i_2 i_3 i_4} = {1\over 120}[n(n^2+1) d^{(4)}_{(i_1 i_2 i_3 i_4)}
-2(n^2-4)\delta_{(i_1 i_2}\delta_{i_3 i_4)}] \quad ,
\label{seVIxxxi}
\ee
and also
\be
t_{i_1 i_2 i_3 i_4 i_5} = \lambda(n) [n(n^2+5) d^{(5)}_{(i_1 i_2 i_3 i_4 i_5)}
-2(3n^2-20) d_{(i_1 i_2 i_3} \delta_{i_4 i_5)}]
\quad,
\label{seVIxxxii}
\ee
by identifying $t_{i_1 i_2 i_3 i_4 i_5}$ as the traceless linear combination
of the appropriate dim${\cal V}^{(5)}=2$ tensors. The factor $\lambda(n)$
has not been determined explicitly: the evaluation
of (\ref{seVIxxxi}) is already time consuming.
These results apply to generic $n$, \ie\ for $n$ such that
dim${\cal V}^{(4)}=2=$dim${\cal V}^{(5)}$.
It is a valuable check on the computations to insert $n=3$ into
(\ref{seVIxxxi}) and $n=4$ into (\ref{seVIxxxi}) and (\ref{seVIxxxii}), and
employ (\ref{seVIiii}), (\ref{seVIiv}) and (\ref{seVIv}) to obtain

\begin{eqnarray}
& \mbox{for\ $n=3$:} &
t_{i_1 i_2 i_3 i_4}
=
{1\over 4} [d^{(4)}_{(i_1 i_2 i_3 i_4)} -
{1\over 3}\delta_{(i_1 i_2}\delta_{i_3 i_4)}] = 0
\quad;
\label{seVIxxxiii}
\\[0.3cm]
&\mbox{for\ $n=3$:} &
t_{i_1 i_2 i_3 i_4 i_5}
=
42 \lambda(3) [d^{(5)}_{(i_1 i_2 i_3 i_4 i_5)} -
{1\over 3} d_{(i_1 i_2 i_3} \delta_{i_4 i_5)}] = 0
\quad;
\label{seVIxxxiv}
\\[0.3cm]
&\mbox{for\ $n=4$:}&
t_{i_1 i_2 i_3 i_4 i_5}
=
84 \lambda(4) [d^{(5)}_{(i_1 i_2 i_3 i_4 i_5)} -
{2\over 3} d_{(i_1 i_2 i_3} \delta_{i_4 i_5)}] = 0
\quad.
\label{seVIxxxv}
\end{eqnarray}
These results exhibit the crucial property of the $t$-tensors. They are in
one-to-one correspondence with the $\Omega$ tensors and hence, as we discuss
below, provide an ideal way (free from problems associated with $d$-tensors
noted above) to discuss Casimir operators.
For low $n$, for which only ($2m-1$)-cocycles with
$2m-1\le 2n-1$ can be
non-zero, the generic results for $t^{(m)}$-tensors
(\ref{seVIxxix})-(\ref{seVIxxxii}) collapse according to
(\ref{seVIxxxiii})-(\ref{seVIxxxv}), as they must when $m>n$,
to give vanishing tensors.

An alternative way to see the same mechanism at work stems from
calculation of the numbers $K^{(m)}(n)$ of (\ref{seVIxxviii}). We
have
\be
{ K^{(m)}=0 \quad m>n \quad,}
\label{zeroescal}
\ee
so that $su(n)$ only defines $(n-1)$ non-zero scalars. By direct
computation based on (\ref{seVIxxix})-(\ref{seVIxxxii}) (using
$d_{ijk}d_{ijk}=(n^2-1)(n^2-4)/n$ [{\it cf.} eq. (\ref{123foldTrs})]
and similar relations, see the Appendix) we get for $m=2,3$ and $4$,
with some work in the last case

\begin{eqnarray}
K^{(2)}(n) & = & n^2(n^2-1)
\label{seVIxxxxa}
\\[0.3cm]
\displaystyle
K^{(3)}(n) & = & -{1\over 144} n^3 (n^2-1) (n^2-4)
\label{seVIxxxxb}
\\[0.3cm]
\displaystyle
K^{(4)}(n) & = & \left({1\over 120}\right)^2{2\over 3}
n^2 (n^2+1) (n^2-1) (n^2-4) (n^2-9)
\label{seVIxxxxc}
\end{eqnarray}
and likewise from (\ref{seVIxxxii})

\be
K^{(5)}(n) = [\lambda(n)]^2 {n\over 3} (n^2+5) \prod_{l=1}^4 (n^2-l^2)\quad.
\label{seVIxxxxi}
\ee
\\
Hence, it follows that
\be
\begin{array}{lllllllll}
K^{(3)}(2)& = & K^{(4)}(2) &=& K^{(5)}(2) &= & \dots &= & 0
\\[0.3cm]
&& K^{(4)}(3) &=& K^{(5)}(3) &= & \dots &= & 0
\\[0.3cm]
&&& & K^{(5)}(4) & = & \dots & = & 0
\end{array}
\label{zeroscalars}
\ee
It is now plausible to conclude that the pattern persists for all
$su(n)$ algebras, and that with the required modifications it remains
true for other classical families.

For low rank groups, for which cocycles above a certain order cannot
appear, the corresponding possibly non-zero scalars $K^{(m)}$ that can
be formed can be seen to vanish identically, as expected since the
$t$-tensors used to define them also do so.

\subsection{Relations for the Casimir operators ${\cal C}'$}

Corollary \ref{cor3.2} exhibits the one-to-one correspondence among
the $t^{(m)}$ tensors and the Casimir invariants ${\cal C}'^{(m)}$.
The properties of the $t$ tensors in Sec. \ref{suse6.2} show now that
we do not get primitive Casimirs of undesired order, since
(\ref{seVIxxxiii}), (\ref{seVIxxxiv}), for instance, yield ${\cal
  C}'^{(4)}=0={\cal C}'^{(5)}$.  This is of course needed to make
sense: there are no $\Omega^{(7)}$, $\Omega^{(9)}$ cocycles for
$su(3)$ since there are only $l$ primitive Casimirs and cocycles for
$su(l+1)$. Similarly (\ref{seVIxxxv}) shows that ${\cal C}'^{(5)}=0$
for $su(4)$ since there is no $\Omega^{(9)}$ for this algebra.

\begin{example}
For $su(n)$ ($n>2$), eq. (\ref{seIIxiii}) gives

\be
{\cal C'}^{(3)} = [\Omega^{(5)}]^{i_1 i_2 i_3 i_4 i_5}
X_{i_1} X_{i_2} X_{i_3} X_{i_4} X_{i_5} = {n^2 \over 12}k^{abc}X_aX_bX_c
={in^2\over 48}d^{abc} X_a X_b X_c \quad,
\label{exam6.6ii}
\ee
as expected from (\ref{nuevocas}) and (\ref{seVIxxx}).
Similarly, if we compute the 4th-order Casimir from (\ref{seIIxiii}), we
obtain
\be
\begin{array}{l}
{\cal C'}^{(4)}=
[\Omega^{(7)}]^{i_1 i_2 i_3 i_4 i_5 i_6 i_7} X_{i_1} X_{i_2} X_{i_3}
X_{i_4} X_{i_5} X_{i_6} X_{i_7}
\\[0.3cm]
\quad\quad
=
\displaystyle
{1 \over 120\cdot 8}(n(n^2+1) {\cal C}^{(4)} - 2(n^2-4)
\delta^{(i_1 i_2}\delta^{i_3 i_4)} X_{i_1} X_{i_2} X_{i_3} X_{i_4})
\\[0.3cm]\quad\quad
=
\displaystyle
{1 \over 120\cdot 8}(n(n^2+1) {\cal C}^{(4)} - 2(n^2-4)([{\cal C}^{(2)}]^2
+{n\over 6}{\cal C}^{(2)}))
\quad.\end{array}
\label{exam6.6iv}
\ee
If we set $n=3$ and use (\ref{seVIvii}) above we get
\be
{\cal C'}^{(4)}=
{1 \over 12\cdot 8}(3 {\cal C}^{(4)} -
[{\cal C}^{(2)}]^2 - {1\over 2}{\cal C}^{(2)})=0\quad.
\label{cprimzero}
\ee
Thus, with the ${\cal C'}$ family, we do not obtain Casimir
operators beyond the order of the higher invariant primitive
polynomial of the algebra.  Also we have ${\cal
  C'}^{(2)}=f^{abc}X_aX_bX_c=(1/2)f^{abc}f_{abd}X^d X_c = (n/2)X^d X_d
$. Recall that the fourth-order Casimir ${\cal C}^{(4)}$ is defined by
${\cal C}^{(4)}= d^{\rho(i_1 i_2} d_\rho ^{i_3) i_4} X_{i_1} X_{i_2}
X_{i_3} X_{i_4}$ (see (\ref{seVIvi})).
\end{example}

We may conclude by stating that the $l$ tensors $t$ introduced by
Lemma \ref{lem3.2}
are in a rather privileged position due to their full `tracelessness', eq.
(\ref{seIIxiia}). In particular, eq. (\ref{seIIxiii}) provides a definition
of the primitive Casimirs ${\cal C}'$ which does not contain
non-primitive terms
and which hence gives zero, as it should, when their order goes beyond
the maximum $m_l$ order permitted.

\subsection{Some technical remarks}

  The crucial results (\ref{seVIxxxi}) and (\ref{seVIxxxii}),
which provide explicit expressions for the fourth and fifth-order invariant
symmetric $su(n)$-tensors require, for their derivation, a series of
expressions involving properties of the $f$ and $d$ tensors of $su(n)$
which at present are not available in the literature. The same applies to
the calculations (\ref{seVIxxxxa})-(\ref{seVIxxxxi}) for the canonical $su(n)$
scalars
$K^{(n)}$, on which we have based some generalisations. In addition to
early work \cite{MSW} containing a modest amount of generic $su(n)$ material,
we have used \cite{KR} which contains a wide class of identities for $f$
and $d$ tensors. To proceed further here, we make use of the fact that an
arbitrary symmetric $SU(n)$-invariant tensor may be expanded in terms of
a basis of ${\cal V}^{(m)}$ as explained in Subsec.\ref{suse6.1}
(see also \cite{SUDBERY,RS} for low values of $m$).
We relegate the detailed treatment to an appendix.

\section{Recurrence relations for primitive cocycles}
\label{invsec7}

Consider first the case of $su(n)$.
In the defining representation, the unit matrix and the $T_i$'s
span a basis for the space of hermitian $n\times n$ matrices.
This means that, since the symmetrised product of hermitian matrices is also
a hermitian matrix, we can write
\be
\{T_{i_1},\dots, T_{i_m}\}\propto {\tilde k}_{i_1\dots i_m}^{\sigma}
T_\sigma + \hat k_{i_1\dots i_m} I
\label{symsun}
\ee
where ${\tilde k}_{i_1\dots i_m}^\sigma$ and $\hat
k_{i_1\dots i_m}$ are invariant symmetric polynomials (of order
$(m+1)$ and $m$) that can be related to the $k$ tensors defined by
(\ref{seIi}).  However, to give recurrence relations we want to use
here the invariant symmetric tensors $d$ (\cite{SUDBERY}; see also
\cite{AK}) defined by eq.  (\ref{VIai})\footnote{Other expressions
  such as (\ref{seIi}) could be used here, since the non-primitive
  parts in (\ref{seIi}) do not contribute to the cocycle definition
  due to Corollary \ref{cor3.1}.}.

The $d^{(m)}$ polynomials are not traceless, and hence differ from
those of (\ref{seIIxii}). Let us use them to define cocycles.  Since
the structure constants $C_{ij}^k$ themselves provide a three-cocycle,
the five-cocycle may be rewritten in the form

\be
{\Omega^{(5)}_{i_1 i_2 i_3 i_4 i_5}}={1\over {5!}}
\epsilon_{i_1 i_2 i_3 i_4 i_5}^{j_1 j_2 j_3 j_4 j_5}
{\Omega^{(3)}_{j_1 j_2}}^{l_1}_\cdot
C^{l_2}_{j_3 j_4} {d_{l_1 l_2 j_5}}
\quad.
\label{seIIIi}
\ee
The next case may be treated similarly.
The coordinates of $\Omega^{(7)}$ are given by
\be
{\Omega^{(7)}_{i_1 i_2 i_3 i_4 i_5 i_6 i_7}}={1\over {7!}}
\epsilon_{i_1 i_2 i_3 i_4 i_5 i_6 i_7}^{j_1 j_2 j_3 j_4 j_5 j_6 j_7}
C^{l_1}_{j_1 j_2} C^{l_2}_{j_3 j_4} C^{l_3}_{j_5 j_6}
d_{l_1 l_2 l_3 j_7}^{(4)}
\label{seIIIii}
\ee
Now, although
\be
d_{l_1 l_2 l_3 j_7}^{(4)}=
{d_{l_1 l_2}}_\cdot^s d_{s l_3 j_7}
\quad.
\label{seIIIiv}
\ee
(see (\ref{VIai})) is not fully symmetric, the extra skewsymmetry
in (\ref{seIIIii}) (cf. (\ref{seIIiv}) and (\ref{seIIv})) permits us
to use (\ref{seIIIiv}) in (\ref{seIIIii}) without having to
symmetrise. \footnote{Note that, in the above examples, it is not
necessary to introduce a five [seven] index $\epsilon$ tensor in the
$r.h.s.$; instead, skewsymmetry in $i_2 i_3 i_4$ [$i_1...i_6$] will
suffice in (\ref{seIIIi}) [(\ref{seIIIv})] because
$d_{(ijk)}=d_{ijk}$ [$d_{(ijkl)}^{(4)}=d_{(ijk)l}^{(4)}$]. However,
for the fifth-order polynomial $d_{ijklm}^{(5)}$ we cannot use such
a simplification (although in the symmetrisation there are fewer than
$5!$ different terms). In this case (and for higher order
$d$-polynomials) we need the complete ($2m-1$)-th order $\epsilon$
tensor, eq. (\ref{seIIiv}), to remove the non-symmetric parts in
(\ref{VIai}).} This means that (\ref{seIIIii}) may be rewritten as

\be
\begin{array}{rl}
\displaystyle
{\Omega^{(7)}_{i_1 i_2 i_3 i_4 i_5 i_6 i_7}} & =
\displaystyle
{1\over {7!}}
\epsilon_{i_1 i_2 i_3 i_4 i_5 i_6 i_7}^{j_1 j_2 j_3 j_4 j_5 j_6 j_7}
C^{l_1}_{j_1 j_2} C^{l_2}_{j_3 j_4}
{d_{l_1 l_2}}_\cdot^s
C^{l_3}_{j_5 j_6} d_{s l_3 j_7}
\\[0.3cm]
&
\displaystyle
= {1\over 3!} {1\over {7!}}
\epsilon_{i_1 i_2 i_3 i_4 i_5 i_6 i_7}^{j_1 k_2 k_3 k_4 j_5 j_6 j_7}
\epsilon_{k_2 k_3 k_4}^{j_2 j_3 j_4}
C^{l_1}_{j_1 j_2} C^{l_2}_{j_3 j_4}
{d_{l_1 l_2}}_\cdot^s
C^{l_3}_{j_5 j_6} d_{s l_3 j_7}
\\[0.3cm]
&
\displaystyle
= {1\over {7!}}
\epsilon_{i_1 i_2 i_3 i_4 i_5 i_6 i_7}^{j_1 k_2 k_3 k_4 j_5 j_6 j_7}
{\Omega^{(5)}_{j_1 k_2 k_3 k_4}}^s_\cdot
C^{l_3}_{j_5 j_6} d_{s l_3 j_7}
\quad.
\end{array}
\label{seIIIv}
\ee
Extending the procedure to an arbitrary $\Omega^{(2m-1)}$ cocycle we now get
the following

\begin{lemma}
({\it Recurrence relation for $su(n)$ primitive cocycles})
\\
Given a $(2(m-1)-1)$-cocycle of $su(n)$, the coordinates of the next
$(2m-1)$-cocycle are obtained from $\Omega^{(2(m-1)-1)}$
by the formula

\be
\Omega^{(2m-1)}_{i_1 \dots i_{2m-1}} = {1 \over {(2m-1)!}}
\epsilon^{j_1 \dots j_{2m-1}}_{i_1 \dots i_{2m-1}}
{\Omega^{(2[m-1]-1)}_{j_1 \dots j_{2m-4}}}^s_\cdot
C^{l}_{j_{2m-3} j_{2m-2}} d_{s l j_{2m-1}}\quad.
\label{seIIIvi}
\ee
\end{lemma}

Clearly, other relations may be found using again (\ref{seIIIvi}) to express
the lower order cocycle $\Omega^{(2[m-1]-1)}$ in terms of
$\Omega^{(2[m-2]-1)}$, etc.

\medskip
For the symplectic and orthogonal algebras $B_l,C_l,D_l$ (setting aside for
$D_l$ the case of the polynomial of order $m_l=l$ related to the Pfaffian) the
primitive symmetric invariant polynomials of order $2,4,\dots,2l$ ($B_l,C_l$)
and $2,4\dots,(2l-2)$ ($D_l$) may be constructed by means of (\ref{hfamily}),
and they lead to primitive cocycles of orders $3,7,\dots,(4l-1)$ ($B_l,C_l$)
and of order $3,7,\dots,(4l-5)$ ($D_l$).
Thus, the first recurrence relation starts for
$\Omega^{(11)}$, which is written as

\be
\Omega^{(11)}_{i_1 \dots i_{11}} = {1\over {11!}}
\epsilon_{i_1 \dots i_{11}}^{j_1 \dots j_{11}}
C^{l_1}_{j_1 j_2} \dots C^{l_5}_{j_9 j_{10}} v_{l_1\dots l_5 j_{11}}
\label{seIIIvii}
\ee
or, using (\ref{hfamily})
\be
\begin{array}{rl}
\Omega^{(11)}_{i_1 \dots i_{11}} =
&
\displaystyle
{1\over {11!}}
\epsilon_{i_1 \dots i_{11}}^{j_1 \dots j_{11}}
C^{l_1}_{j_1 j_2} C^{l_2}_{j_3 j_4} C^{l_3}_{j_5 j_6} C^{l_4}_{j_7 j_8}
C^{l_5}_{j_9 j_{10}}  {v_{l_1 l_2 l_3}}_\cdot^s v_{s l_4 l_5 j_{11}}
\\[0.3cm]
=
&
\displaystyle
{1\over 5!} {1\over {11!}}
\epsilon_{i_1 \dots \dots \dots \dots \dots i_{11}}
^{j_1 k_2 \dots k_6 j_7 \dots j_{11}}
\epsilon_{k_2 \dots k_{6}}^{j_2 \dots j_{6}}
C^{l_1}_{j_1 j_2} C^{l_2}_{j_3 j_4} C^{l_3}_{j_5 j_6}
{v_{l_1 l_2 l_3}}_\cdot^s
C^{l_4}_{j_7 j_8} C^{l_5}_{j_9 j_{10}}  v_{s l_4 l_5 j_{11}}
\\[0.3cm]
=
&
\displaystyle
{1\over {11!}}
\epsilon_{i_1 \dots i_{11}}^{j_1 \dots j_{11}}
{\Omega^{(7)}_{j_1 \dots j_6}}_\cdot^s
C^{l_4}_{j_7 j_8}C^{l_5}_{j_9 j_{10}} v_{s l_4 l_5 j_{11}}
\quad,
\end{array}
\label{seIIIviii}
\ee
which expresses $\Omega^{(11)}$ in terms of $\Omega^{(7)}$ and the
fourth-order polynomial.
This leads to the following recurrence relation

\begin{lemma}
({\it Recurrence relation for $so(2l+1),sp(l),so(2l)$})
\\
Let $\Omega^{(4p-1)}$ be a $(4p-1)$-cocycle for $so(2l+1)$, $sp(l)$
$(p=1,\dots,l)$, $so(2l)$ $(p=1,\dots,(l-1))$.
Then,
\be
\Omega^{(4p-1)}_{i_1 \dots i_{4p-1}} =
{1\over (4p-1)!}
\epsilon_{i_1 \dots i_{4p-1}}^{j_1 \dots j_{4p-1}}
{\Omega^{(4[p-1]-1)}_{j_1 \dots j_{4p-6}}}_\cdot^s
C^{l_1}_{j_{4p-5} j_{4p-4}}C^{l_2}_{j_{4p-3} j_{4p-2}} v_{s l_1 l_2 j_{4p-1}}
\,.
\label{seIIIix}
\ee
\end{lemma}

\section{Duality relations for skewsymmetric primitive tensors}
\label{invsec8}

The interpretation of the primitive cocycles as closed forms on the group
manifold of a simple compact group $G$ provides another intuitive way to obtain
additional relations among them.
Consider the case of $su(2)$.
The identification of the $su(2)$-three-cocycle with the closed de Rham
three-form on
$SU(2)\sim S^3$ tells us that this form is (up to a constant)
the volume form on the $SU(2)$ group manifold.
It is known \cite{CAR,PON,HOPF,HOD,CE,SAMEL,BOREL,BOTT,LJB} that,
from the point of view of real homology, the compact
groups behave as  `products' of spheres
of odd dimension $(2m_i-1)$, $i=1,\dots,l$.
As a result,
\be
\Omega=
\Omega^{(2m_1-1)}\wedge\dots\wedge \Omega^{(2m_l-1)}
\label{seIVi}
\ee
is proportional to the volume form on the group manifold and, indeed,
$\sum_{i=1}^l (2m_i-1) = r = \mbox{dim} G$ for all simple groups.
For instance, if $G=SU(3)$
\be
\Omega(g)=
\Omega^{(3)}(g)\wedge \Omega^{(5)}(g) \propto
C_{i_1 i_2 i_3}
\Omega^{(5)}_{i_4 i_5 i_6 i_7 i_8}
\omega^{i_1}(g)\wedge\dots\wedge \omega^{i_8}(g)
\label{seIVii}
\ee
is proportional to
$\omega^1(g)\wedge\dots\wedge \omega^8(g)$, the volume element on $SU(3)$.

Algebraically $(\omega^i(g)\to \omega^i)$,
we may look at (\ref{seIVii}) as the wedge product of two
skewsymmetric tensors defined on a vector space $V$ of dimension $r$.
If we now endow $V$ with a metric (the unit metric for $\g$ compact)
we may introduce the Hodge $*$ operator.
Then, the scalar product of two skewsymmetric tensors
$\alpha={1\over q!}\alpha_{i_1\dots i_q}\omega^{(i_1)}\wedge\dots
\wedge\omega^{(i_q)}$ and $\beta$ of order $q$ is given by
\be
<\alpha,\beta>=\alpha\wedge (* \beta)={1\over q!} \alpha_{i_1\dots i_q}
\beta^{i_1\dots i_q}\omega^1\wedge\dots\wedge\omega^r \quad .
\label{seIViii}
\ee
Consider the simplest $su(2)$ case. There is only one cocycle
($cf$. (\ref{seIIiii}))
\be
\Omega^{(3)}={1\over 3!}\Omega_{i_1 i_2 i_3}
\omega^{i_1}\wedge \omega^{i_2}\wedge \omega^{i_3}
\quad.
\label{seIViv}
\ee
We fix the normalisation by demanding that
\be
<\Omega^{(3)},\Omega^{(3)}>=\Omega^{(3)}\wedge (* \Omega^{(3)}) =
{1\over 3!}\Omega_{i_1 i_2 i_3} \Omega^{i_1 i_2 i_3}
\omega^1\wedge \omega^{2}\wedge \omega^{3}
= \omega^1\wedge \omega^{2}\wedge \omega^{3}
\label{seIVva}
\ee
($\omega^1(g)\wedge\omega^2(g)\wedge\omega^3(g)$ is the volume element
on $SU(2)$), \ie,
\be
{1\over 3!}\Omega_{i_1 i_2 i_3} \Omega^{i_1 i_2 i_3} = 1
\quad,
\label{seIVv}
\ee
which is trivially satisfied since
$\Omega_{i_1 i_2 i_3}=\epsilon_{i_1 i_2 i_3}$ for $su(2)$.
Let now $G=SU(3)$, and let the five cocycle be expressed as by
\be
\Omega^{(5)}={1\over 5!}\Omega_{i_1 i_2 i_3 i_4 i_5}
\omega^{i_1}\wedge \omega^{i_2}\wedge \omega^{i_3} \wedge
\omega^{i_4}\wedge \omega^{i_5}
\quad.
\label{seIVvi}
\ee
We now fix now the normalisations of $\Omega^{(3)}$ and $\Omega^{(5)}$ by
requiring that
\be
\Omega^{(3)}\wedge (*\Omega^{(3)}) =\omega^1\wedge\dots\wedge\omega^8=
\Omega^{(5)}\wedge (*\Omega^{(5)})
\label{seIVvii}
\ee
This gives the previous relation for the coordinates of $\Omega^{(3)}$
and a similar one for those of $\Omega^{(5)}$.
Up to an irrelevant sign (which is a minus
sign for even $r$, as is the case of
$SU(3)$, since $*^2=(-1)^{q(r-q)}$ where $q$ is the order (always odd) of
the cocycle) we may write
\be
\Omega^{(5)}=* \Omega^{(3)}
\label{seIVix}
\ee
(for a positive definite metric $<\alpha,\beta>=<*\alpha,*\beta>$)
and hence (with $r$=8)
\be
\Omega^{(5)}={1\over (r-3)!} {1\over 3!}
\delta^{i_1 j_1} \delta^{i_2 j_2} \delta^{i_3 j_3}
\epsilon_{j_1 j_2 j_3 l_1 l_2 l_3 l_4 l_5}
\Omega^{(3)}_{i_1 i_2 i_3}
\omega^{l_1}\wedge \omega^{l_2}\wedge \omega^{l_3} \wedge
\omega^{l_4}\wedge \omega^{l_5}
\label{seIVx}
\ee
\ie,
\be
\Omega^{(5)}_{l_1 l_2 l_3 l_4 l_5} =
{1\over 3!} \epsilon_{j_1 j_2 j_3 l_1 l_2 l_3 l_4 l_5}
{\Omega^{(3)}}^{j_1 j_2 j_3}
\quad.
\label{seIVxi}
\ee

The previous arguments are not restricted to $su(n)$ nor to the case of two
cocycles. In general we have that relation (\ref{seIVi}) holds and, as a
result, we find up to irrelevant signs a whole series of duality
relations among cocycles:
\be
\begin{array}{l}
\displaystyle
\Omega^{(2m_i-1)}=
*(\Omega^{(2m_1-1)}\wedge\dots \widehat{\Omega^{(2m_i-1)}}\wedge\dots
\Omega^{(2m_l-1)})
\\[0.3cm]
\displaystyle
\Omega^{(2m_i-1)}\wedge \Omega^{(2m_j-1)}
=
*(\Omega^{(2m_1-1)}\wedge\dots \widehat{\Omega^{(2m_i-1)}}\wedge\dots
\widehat{\Omega^{(2m_j-1)}}\wedge\dots\Omega^{(2m_l-1)})
\\[0.3cm]
\dots\dots\dots
\end{array}
\label{seIVxii}
\ee
where $1\le i,j,\dots\le l$, etc.
In general, the normalisation of the $(2m_i-1)$ cocycles may be introduced
by requiring that
\be
<\Omega^{(2m_i-1)},\Omega^{(2m_i-1)}>
=\Omega^{(2m_i-1)} \wedge (*\Omega^{(2m_i-1)})
=\omega^1\wedge\dots\wedge\omega^r \quad ,
\label{seIVxiii}
\ee
the volume element on $G$.
\begin{example}
The three- and five-cocycles for $su(3)$, given in co-ordinates in
tables \ref{table1} and \ref{table2} respectively, satisfy the duality
relation

\begin{equation}
\Omega_{i_1 i_2 i_3 i_4 i_5} = \frac{1}{3! 2 \sqrt{3}}
\epsilon_{i_1 i_2 i_3 i_4 i_5 i_6 i_7 i_8} f^{i_6 i_7 i_8}
\label{seVIIIi}
\end{equation}
\end{example}

\begin{example}
For $su(4)$, we use the definitions of the three- and five-cocycles of
tables \ref{table4} and \ref{table6} and the expression (\ref{seIIIv})
of the seven-cocycle $\Omega^{(7)}$ to obtain the following relationship:

\be
15\sqrt{2} \Omega^{(7)}_{i_1 i_2 i_3 i_4 i_5 i_6 i_7} = \frac{1}{5!3!}
\epsilon_{i_1 i_2 i_3 i_4 i_5 i_6 i_7 j_1 j_2 j_3 j_4 j_5 k_1 k_2 k_3}
\Omega^{(5)}_{j_1 j_2 j_3 j_4 j_5} \Omega^{(3)}_{k_1 k_2 k_3}
\label{ochodos}
\ee
\end{example}

These relations have been computed using MAPLE and provide a further
check of the cocycle Tables in Sec. \ref{cocycles}.

\subsection*{Acknowledgements}

This research has been partially supported by the CICYT and the DGICYES,
Spain.
J.A. and J.C.P.B. wish to acknowledge the kind hospitality extended to them
at DAMTP and J.C.P.B. wishes to thank the Spanish Ministry of Education and
Culture and the CSIC for an FPI grant.

\setcounter{section}{0}
\renewcommand{\thesection}{\appendixname~\Alph{section}:}
\renewcommand{\theequation}{\Alph{section}.\arabic{equation}}

\section{Traces of products of $su(n)$ $D$ and $F$ matrices}

We define the hermitian $D$ and antihermitian $F$ (adjoint) traceless
matrices for arbitrary $su(n)$ by

\begin{equation}
  (F_a)_{bc} = f_{bac} \quad , \quad (D_a)_{bc} = d_{abc}\quad ,\quad
a,b,c=1,...,n^2-1\quad,
\label{ap1}
\end{equation}
intending to present the identities that we require involving $d$ and
    $f$ tensors of $SU(n)$ in terms of traces of products of $D$ and
    $F$ matrices. All the 2 and 3-fold traces have been known for a
    long time \cite{KR,MSW}. Explicitly,
\begin{equation}
  \begin{array}{lll}
\displaystyle       \mbox{Tr} F_a F_b = -n \delta_{ab} ,
&
\displaystyle   \mbox{Tr} F_a D_b = 0 ,
&
\displaystyle   \mbox{Tr} D_a D_b =
      \frac{n^2-4}{n} \delta_{ab} ,
\\[0.3cm]
\displaystyle
\mbox{Tr} F_a F_b F_c = -\frac{n}{2} f_{abc} , &
\displaystyle
\mbox{Tr} F_a F_b D_c =
      -\frac{n}{2} d_{abc} , &
\\[0.3cm]
\displaystyle
\mbox{Tr} F_a D_b D_c = \frac{n^2-4}{2n} f_{abc} ,
&
\displaystyle   \mbox{Tr} D_a D_b D_c =      \frac{n^2-12}{2n} d_{abc} \quad .
&
\end{array}
  \label{123foldTrs}
\end{equation}
The methods of \cite{KR} yield only
expressions for such four-fold traces as

\begin{equation}
  \mbox{Tr} F_a F_b F_c D_d \quad , \quad \mbox{Tr} F_a D_b D_c D_d ,
\label{ap2}
\end{equation}
as well as all others that follow from these which involve an odd
number of $F$ and $D$ matrices. To proceed further (to the evaluation
of the traces of all four-fold products of even numbers of $D$ and $F$
matrices), we begin by treating

\begin{equation}
  \mbox{Tr} F_{(a} F_b F_c F_{d)} \quad \mbox{and} \quad \mbox{Tr} D_{(a} D_b
D_c D_{d)}
\quad .
\label{4foldsTrs}
\end{equation}
Once this is done, $\mbox{Tr} F_a F_b F_c F_d$, $\mbox{Tr} D_a D_b D_c
D_d$, $\mbox{Tr} F_a F_b D_c D_d$ (and other similar traces) can be
calculated by means of further elementary procedures. We list results
valid for arbitrary $su(n)$.

\begin{eqnarray}
  \mbox{Tr} F_a F_b F_c F_d & = & \delta_{ab} \delta_{cd} + \delta_{ad}
    \delta_{bc} + \frac{n}{4} (d_{abx} d_{cdx} + d_{adx} d_{bcx} -
    d_{acx} d_{bdx}) , \label{trFFFF} \\[0.3cm]
  \mbox{Tr} F_a F_b F_c D_d & = & -\frac{n}{4} d_{abx} f_{cdx} - \frac{n}{4}
    f_{abx} d_{cdx} , \label{trFFFD} \\[00.3cm]
  \mbox{Tr} F_a F_b D_c D_d & = & \frac{4-n^2}{n^2} (\delta_{ab} \delta_{cd}
    - \delta_{ac} \delta_{bd}) + \frac{8-n^2}{4n} (d_{abx} d_{cdx} -
    d_{acx} d_{bdx}) \nonumber \label{trFFDD} \\[0.3cm]
    & & \mbox{} - \frac{n}{4} d_{adx} d_{bcx} , \\[0.3cm]
  \mbox{Tr} F_a D_b F_c D_d & = & \frac{n}{4} (d_{acx} d_{bdx} - d_{adx}
d_{bcx})
    - \frac{n}{4} d_{abx} d_{cdx} , \label{trFDFD} \\[0.3cm]
  \mbox{Tr} F_a D_b D_c D_d & = & \frac{n^2-12}{4n} f_{abx} d_{cdx} +
    \frac{n}{4} d_{abx} f_{cdx} \nonumber \label{trFDDD} \\[0.3cm]
    & & \mbox{} + \frac{1}{n} (f_{adx} d_{bcx} - f_{acx} d_{bdx}) , \\[0.3cm]
  \mbox{Tr} D_a D_b D_c D_d & = & \frac{n^2-4}{n^2} (\delta_{ab} \delta_{cd} +
    \delta_{ad} \delta_{bc}) - \frac{n}{4} d_{acx} d_{bdx} \nonumber \\[0.3cm]
  & & \mbox{} + \frac{n^2-16}{4n} (d_{abx} d_{cdx} + d_{adx} d_{bcx})
    . \label{trDDDD}
\end{eqnarray}
Now we illustrate the method of derivation of the above results and
perform a variety of checks of their correctness. In the framework
of Sec. \ref{invsym}, we set out from an expansion of an arbitrary
totally symmetric fourth-rank tensor in ${\cal V}^{(4)}$
and write
\begin{equation}
  \mbox{Tr} F_{(a} F_b F_c F_{d)} =
A d^{(4)}_{(abcd)} + B \delta_{(ab} \delta_{cd)}
\quad .
\label{ap3}
\end{equation}
Contracting both sides with $\delta_{ab}$ and $d_{abe}$ in turn and
using results such as (\ref{123foldTrs}) allows us easily to find $A =
n/4$, $B = 2$. Various checks on, {\it e.g.}, (\ref{trFFFF}) and its
consequences now exist. Firstly, in \cite{KR} we find identities for
contractions of (\ref{trFFFF}), (\ref{trFFDD}) and (\ref{trDDDD}) with
$d_{ace}$ and $f_{ace}$. Performing such contractions explicitly on
our expressions for these traces, we find total agreement for all
$SU(n)$. For the case of $SU(3)$ a more elementary derivation of, {\it
e.g.} $\mbox{Tr} F_a F_b F_c F_d$ is available because additional identities
for $SU(3)$ $D$ and $F$ matrices exist as described in \cite{MSW}. One
can thus get the $n = 3$ version of, {\it e.g.} $\mbox{Tr} F_{(a} F_b F_c
F_{d)}$ without using any assumptions about symmetric tensors. In
particular, for $SU(3)$, we have

\begin{eqnarray}
  \mbox{Tr} F_{(a} F_b F_c F_{d)} & = & \frac{9}{4} \delta_{(ab} \delta_{cd)}
\label{ap4.1}    , \\[0.3cm]
  \mbox{Tr} D_{(a} D_b D_c D_{d)} & = & \frac{17}{36} \delta_{(ab} \delta_{cd)}
,
\label{ap4.2}    \\[0.3cm]
  \mbox{Tr} D_a D_b D_c D_d & = & \frac{5}{9} \left( \delta_{ab} \delta_{cd}
    + \delta_{ad} \delta_{bc} \right) - \frac{7}{12} \delta_{(ab}
    \delta_{cd)} - \frac{1}{6} d_{acx} d_{bdx} .
\label{ap4.3}
\end{eqnarray}
Analogously, from
\be
\mbox{Tr} D_{(a} D_b D_c D_{d)} = A'd^{(4)}_{(abcd)} +
B' \delta_{(ab} \delta_{cd)}
\label{ap3a}
\ee
and contracting with $\delta_{ab}$ and $d_{abs}$ we deduce that
$A'=(n^2-32)/4n$ and $B'=2(n^2-4)/n^2$.

Armed with the results (\ref{trFFFF}) to (\ref{trDDDD}) as well as
those of (\ref{123foldTrs}), we can consider five-fold traces. Thus we
postulate (see (\ref{seVIii}))
\begin{equation}
  \mbox{Tr} D_{(a} D_b D_c D_d D_{e)} = A d^{(5)}_{(abcde)}
+ B \delta_{(ab} d_{cde)} \quad,
  \label{5symmDs}
\end{equation}
Contracting this with $\delta_{ab}$ and $d_{abf}$ in turn gives three
linear equations for the coefficients $A$ and $B$; in the latter
construction equating coefficients of $d^{(4)}_{(cde)f} =
d^{(4)}_{(cdef)}$ and $\delta_{(cd} \delta_{e)f} =
\delta_{(cd} \delta_{ef)}$ has actually given rise to two
equations. These three equations are consistent and give
\begin{equation}
  A = \frac{n^2-80}{8n} \quad , \quad B = \frac{3n^2-20}{n^2} \quad .
  \label{AandB}
\end{equation}
The major cancellations that bring (\ref{AandB}) to the form displayed
convince us of the correctness of our results. There are also
other independent
checks available because results for various five-fold traces with no
more than three free indices (vertex corrections, in the diagrammatic
language) are known, which we can reproduce. We also note
that (\ref{5symmDs}) simplifies, not only in the case of $SU(3)$
when (\ref{seVIv}) is used to obtain

\begin{equation}
  \mbox{Tr} D_{(a} D_b D_c D_d D_{e)} = -\frac{5}{24} d_{(abc} \delta_{de)}
\quad ,
\label{ap5}
\end{equation}
but also for $SU(4)$, after use of (\ref{seVIiv}), giving

\begin{equation}
  \mbox{Tr} D_{(a} D_b D_c D_d D_{e)} = \frac{5}{12} d_{(abc} \delta_{de)} .
\label{ap6}
\end{equation}
Note that (\ref{ap5}) and (\ref{ap6}) may be obtained by
contracting the expression

\begin{equation}
  \mbox{Tr} D_{(a} D_b D_c D_d D_{e)} = A'' d_{(abc} \delta_{de)}
\end{equation}
with $\delta_{ab}$; the resulting equation gives $A''=-\frac{5}{24}$
in the $su(3)$ case and $A''=\frac{5}{12}$ for $su(4)$.

Finally, we note that, once (\ref{5symmDs}) has been evaluated and
further five-fold traces obtained from it by elementary procedures, we
have all we need to establish the result

\begin{eqnarray}
  \mbox{Tr} D_{(a} D_b D_c D_d D_e D_{f)} & = & 4 \frac{n^2-4}{n^3}
\delta_{(ab}
    \delta_{cd} \delta_{ef)} + \frac{n^2-192}{16n} {d_{(ab}}^x
    {d_{cd}}^y {d_{ef)}}^z d_{xyz} \nonumber \\[0.3cm]
  & + & \frac{3}{4} \frac{3n^2-64}{n^2} \delta_{(ab} {d_{cd}}^x
    d_{ef)x} + \frac{5n^2+48}{4n^2} d_{(abc} d_{def)} . \label{str6Ds}
\end{eqnarray}
This involves expansion of the left side in terms of a basis
\cite{SUDBERY,RS}, of totally symmetric sixth-rank tensors. Various
contractions give four linear equations for the coefficients involved.
Regarding the expansion (\ref{str6Ds}) we remark that the tensor
${d_{(ab}}^j\,{d^{jk}}_c\, {d^{kl}}_d\,{d^l}_{ef)}$ differs from the
tensor of the second term by $2/n$ times the difference of the tensors
of the fourth minus the first term.

Some further comments are now in order. Many of the procedures followed
above have been guided by graphical ideas such as described in
\cite{CVI}. These simplify complicated expressions involving $d$
and $f$ tensors by reference to closed loops in their graphical
representations. These loops correspond to traces and we set about the
simplification of three line loops with the aid of (\ref{123foldTrs}).
Then we learn how to handle in turn all loops of four and five
internal lines. The calculations associated with (\ref{str6Ds})
are organised by looking at contractions diagrammatically, preferring
to simplify at all times
by identifying loops of as few lines as
possible. It should be noted that many of the identities of this
appendix have been evaluated by different means
including the use of MAPLE. Also. we emphasise that a large number
of checks of our results were made by passing to
subcases for which identities were given in
\cite{KR}.

We turn next to the use of the identities presented in this appendix
in the derivation of results quoted in the body of the
paper. Consider, {\it e.g.} (\ref{seVIxxxi})

\begin{equation}
  t_{p_1 p_2 p_3 p_4} = A d^{(4)}_{(p_1 p_2 p_3 p_4)} + B \delta_{(p_1
    p_2} \delta_{p_3 p_4)} .
\label{ap7}
\end{equation}
Contractions with $\delta_{p_1 p_2}$ and $d_{p_1 p_2 q}$ on the right
are easy to do, as is the contraction of the left side with
$\delta_{p_1 p_2}$. The latter gives zero by direct calculation
as required by Lemma \ref{lem3.2}. The
contraction $t_{p_1 p_2 p_3 p_4} d_{p_1 p_2 q}$ is not governed by any
general argument. To compute it, we set out from (see (\ref{seIIxii})),

\begin{equation}
  t_{pqrg} = \Omega^{(7)}_{abcdefg} f_{abp} f_{cdq} f_{efr} ,
\label{ap8}
\end{equation}
and the cocycles defined using the $d$-family

\begin{eqnarray}
  \Omega^{(7)}_{abcdefg} & = & \Omega^{(5)}_{t[abcd} {f^z}_{ef]}
    d_{tzg} \label{ap9}, \\[0.3cm]
  \Omega^{(5)}_{tabcd} f_{abs} & = & \frac{n}{2} f_{u[cd} d_{t]us} ,
\label{ap10}
\end{eqnarray}
((\ref{seIIIv}), $cf.$ (\ref{seIIIi})) for $SU(n)$. Patient evaluation
of the terms involved can be completed with the aid of four-fold
traces evaluated earlier in this appendix and not (it seems) without
them.  It is clear that the occurrence of $f$-tensors in the cocycle
definitions is what requires us to know how to treat traces involving
$F$ as well as $D$ matrices. Equation (\ref{seVIxxxi}) emerges after
typical and reassuring cancellations. To obtain (\ref{seVIxxxii}) with
$\lambda(n)$ left undetermined, it is necessary only to compute the
ratio of the two scalars occurring in the expansion with respect to the
basis \cite{SUDBERY,RS} of ${\cal V}^{(5)}$ of the left side of
(\ref{seVIxxxii}).  This requires only contraction with $\delta_{p_1
  p_2}$ which is zero.

To prove the result for $K^{(5)}(n)$ in (\ref{seVIxxxxi}), we set out from
(\ref{seVIxxxii}). The only hard part involves

\begin{equation}
  d^{(5)}_{(abcde)} d^{(5)}_{(abcde)}   =   d^{(5)}_{(abcde)}
    d^{(5)}_{abcde}  =   d^{(5)}_{(abcde)} d_{abx} d_{xcy} d_{yde} \quad ,
  \label{d5d5}
\end{equation}
and at worst two equal terms of the fifteen involved lead to a
contracted four-fold trace of $D$ matrices known from \cite{KR}. The
result is
$d^{(5)}_{(abcde)}d^{(5)}_{abcde}={(n^2-4)(n^2-1)\over 15n^3}
(5n^4 - 96 n^2 + 480)$.

\end{document}